\documentclass[
 reprint,
 amsmath,
 pre,
 amssymb,
 nofootinbib,
 aps,
]{revtex4-2}

\usepackage[pdftex]{graphicx}
\usepackage{wasysym}
\usepackage{amsfonts}
\usepackage{ifsym}

\usepackage{amsmath,amssymb,amsfonts,amscd,mathrsfs}
\usepackage{dsfont}

\usepackage{lipsum}
\usepackage{pifont}
\usepackage{graphicx}

\usepackage{tikz-feynman}
 \tikzfeynmanset{compat=1.0.0}

\makeatletter
\newlength{\apb@width}
\newcommand{\autoparbox}[2][c]{\settowidth{\apb@width}{#2}\parbox[#1]{\apb@width}{#2}}

\makeatother

\newcommand{\namedref}[2]{\hyperref[#2]{#1~\ref*{#2}}}

\newcommand{\be}{\begin{equation}}
\newcommand{\ee}{\end{equation}}

\newcommand{\reef}[1]{(\ref{#1})}

\def\bea#1\eea{\begin{align}#1\end{align}}

\renewcommand{\Re}{\mathop{\mathrm{Re}}}
\renewcommand{\Im}{\mathop{\mathrm{Im}}}

\newcommand{\Csphere}{{}^\bullet\kern-1.2pt C}
\newcommand{\Ctorus}{{}^\circ\kern-1.2pt C}

\newcommand{\COMMENT}[1]{}

\newcommand{\neqa}{\nonumber\end{eqnarray}}

\newcommand{\<}{{\langle}}
\renewcommand{\>}{{\rangle}}

\newcommand{\re}{\relax{\rm I\kern-.18em R}}

\def\su2{{SU(2)}}

\def\[{\left[}
\def\]{\right]}

\def\({\left(}
\def\){\right)}
\def\[{\left[}
\def\]{\right]}

\def\<{\langle}
\def\>{\rangle}

\def\i2{\frac{i}{2}}

\def\2F1{\,_2{\rm F}_1}

\usepackage{color}
\usepackage{graphicx}
\usepackage{dcolumn}
\usepackage{bm}
\usepackage{hyperref}

\usepackage{cancel}
\usepackage{ctable}
\usepackage{booktabs}
\newcolumntype{L}[1]{>{\raggedright\let\newline\\\arraybackslash\hspace{0pt}}m{#1}}
\newcolumntype{C}[1]{>{\centering\let\newline\\\arraybackslash\hspace{0pt}}m{#1}}
\newcolumntype{R}[1]{>{\raggedleft\let\newline\\\arraybackslash\hspace{0pt}}m{#1}}

\newcommand{\beq}{\begin{equation}}
\newcommand{\eeq}{\end{equation}}
\newcommand{\beqq}{\begin{equation*}}
\newcommand{\eeqq}{\end{equation*}}
\newcommand\beqa{\begin{eqnarray}}
\newcommand\eeqa{\end{eqnarray}}
\newcommand\beqaa{\begin{eqnarray*}}
\newcommand\eeqaa{\end{eqnarray*}}

\begin{document}


\title{ 
 The  Phases of the Scalar S-Matrix Island 
}

\author{Joan Elias Mir\'o~$^{(a)}$}
\author{Andrea Guerrieri~$^{(b)}$} 
\author{Mehmet Asım Gümüş~$^{(c)}$} 
\affiliation{\vspace{.2cm}    $^{(a)}$~The Abdus Salam ICTP,    Strada Costiera 11, 34135, Trieste, Italy }
\affiliation{\vspace{.2cm}    $^{(b)}$~City St.~George’s, Univ. of London, Northampton Square, EC1V~0HB, London, UK}
\affiliation{\vspace{.2cm}    $^{(c)}$~LAPTh, CNRS et USMB, 9 Chemin de Bellevue, F-74941 Annecy, France }

\begin{abstract}

The two-to-two four-dimensional scattering amplitude of identical scalars obeys rigorous two-sided non-perturbative bounds derived via the modern numerical S-matrix bootstrap. These bounds carve out an allowed region with a rich boundary structure, featuring edges and vertices. In this work we further tighten this region and uncover the physics of its boundary by analyzing the asymptotic Regge behavior of the amplitude and the spectrum of resonances and virtual states. We find that the S-matrices along a given edge exhibit universal behavior, sharply contrasting with that on other edges. This reveals a classification of the boundary into distinct phases, corresponding to different UV mechanisms by which a gapped scalar arises.

\end{abstract}

\pacs{Valid PACS appear here} 

\maketitle



\section{Introduction}

\noindent General principles such as unitarity, analyticity, and crossing symmetry place powerful constraints on physical observables in quantum field theory (QFT). The Bootstrap program exploits these constraints to determine them directly from first principles, without relying on a microscopic Lagrangian. 
The S-matrix Bootstrap \cite{Paulos:2016fap,Paulos:2016but,Paulos:2017fhb,Homrich:2019cbt} applies these ideas directly to scattering processes and has emerged as a robust framework for exploring the landscape of consistent amplitudes with applications ranging from strongly coupled QFTs \cite{Andrea,Bose:2020shm,Bose:2020cod,Hebbar:2020ukp,Albert:2022oes, Karateev:2022jdb,Fernandez:2022kzi,Guerrieri:2023qbg,He:2023lyy,Albert:2023seb,Guerrieri:2024jkn,He:2024nwd} to Quantum Gravity \cite{Guerrieri:2021ivu, Bern:2021ppb, Guerrieri:2022sod, Caron-Huot:2021rmr,Albert:2024yap, Cheung:2024uhn, Bellazzini:2025shd, Beadle:2025cdx, Cheung:2025tbr} and Effective Field Theories \cite{ Bellazzini:2020cot,Tolley:2020gtv, Caron-Huot:2020cmc, Guerrieri:2020bto,Arkani-Hamed:2020blm, Haring:2022sdp,
Acanfora:2023axz,EliasMiro:2023fqi, CarrilloGonzalez:2023cbf, 
Bellazzini:2025bay}.

In two spacetime dimensions, the space of consistent scattering amplitudes is remarkably well understood \cite{Cordova:2018uop,He:2018uxa,Doroud:2018szp,FluxTube,Kruczenski:2020ujw,Guerrieri:2020kcs,EliasMiro:2021nul,Guerrieri:2024ckc,Copetti:2024dcz,Albert:2026fqj}. S-matrices occupy compact regions in spaces of low energy data, with integrable theories -- whose amplitudes obey factorization in addition to the standard bootstrap constraints -- often sitting at cusps \cite{Cordova:2019lot,Cordova:2023wjp,Cordova:2025bah}. In higher dimensions our understanding remains fragmentary. Four-dimensional scalar amplitudes provide a natural laboratory for this problem: they are the simplest setting for numerical exploration, while remaining directly connected to phenomenologically relevant degrees of freedom, from composite states such as mesons to elementary scalars such as the Higgs boson.

\begin{figure}[t]
    \centering
    \includegraphics[scale=0.178]{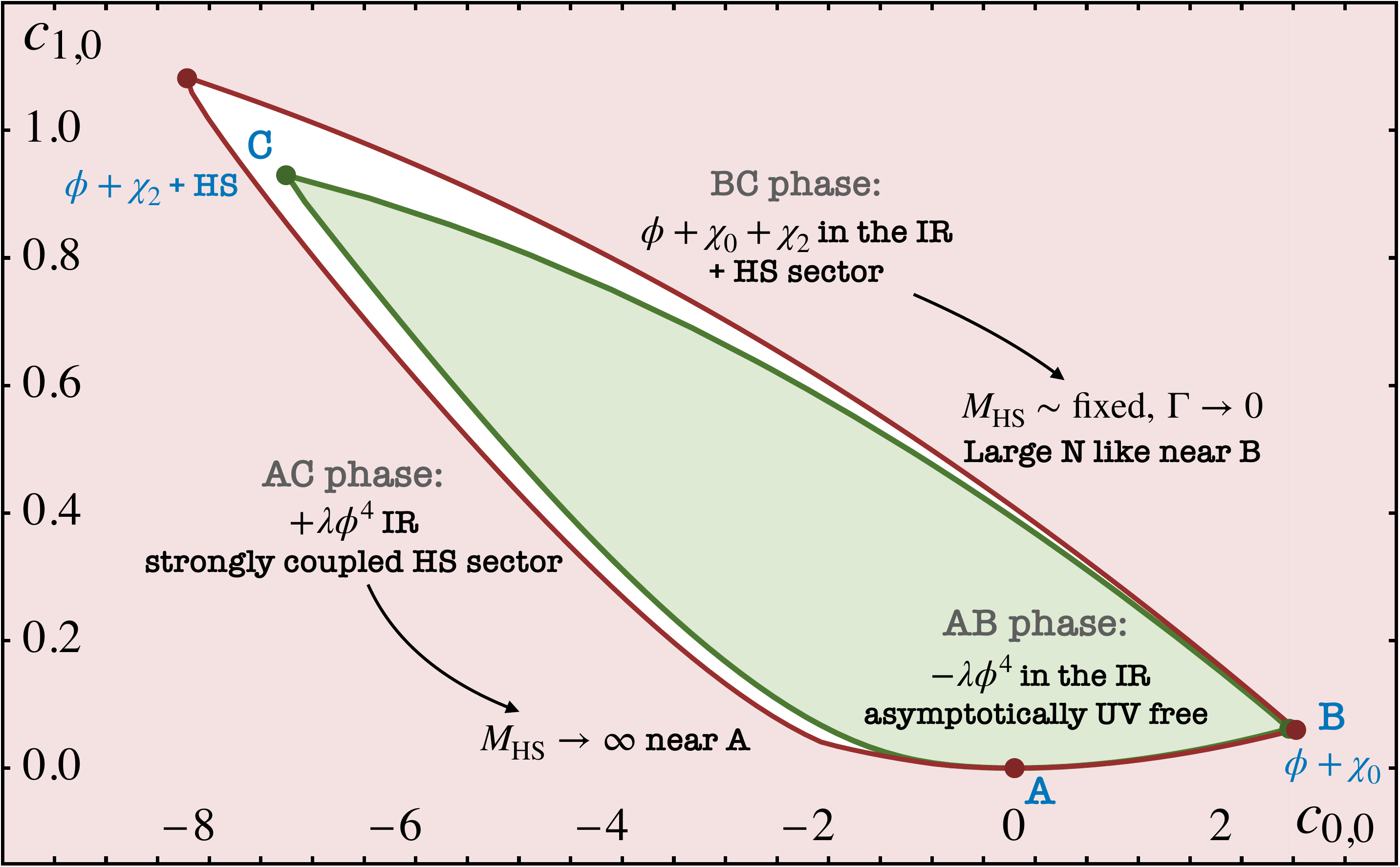}
    \caption{
Projection of the four-dimensional scalar S-matrix space onto the $(c_{0,0},c_{1,0})$ plane. 
The red region is excluded by the dual bootstrap, while the green region is realized by explicit primal amplitudes. 
The free point $A$, 
the cusp $B$ with the scalar-threshold state $\chi_0$, 
and cusp $C$ with the spin-two-threshold state $\chi_2$ are connected by boundary arcs that define three universal phases.
}
    \label{fig:dual_vs_primal}	
\end{figure}

\medskip

In this Letter we study the space of four-dimensional gapped scalar $2\to2$ scattering amplitudes with unit mass $m^2=1$. We characterize amplitudes by their low-energy expansion around the crossing-symmetric point $s=t=u=4/3$,
\begin{equation}
T(s,t)
=32\pi\sum_{n,m} c_{n,m}\,\sigma^n \tau^m ,
\label{eq:low_energy_expansion}
\end{equation}
where $\sigma=\bar s^2+\bar t^2+\bar u^2$, 
$\tau=\bar s\bar t\bar u$, and 
$\bar s=s-\frac{4}{3}$, with cyclic permutations understood. The lowest coefficients define the simplest coordinates on the space of infrared S-matrix data: $c_{0,0}$ controls the quartic scalar coupling, while $c_{1,0}$ governs the leading higher-derivative correction.

\begin{figure*}[t!]
  \centering
  \includegraphics[width=0.435\linewidth]{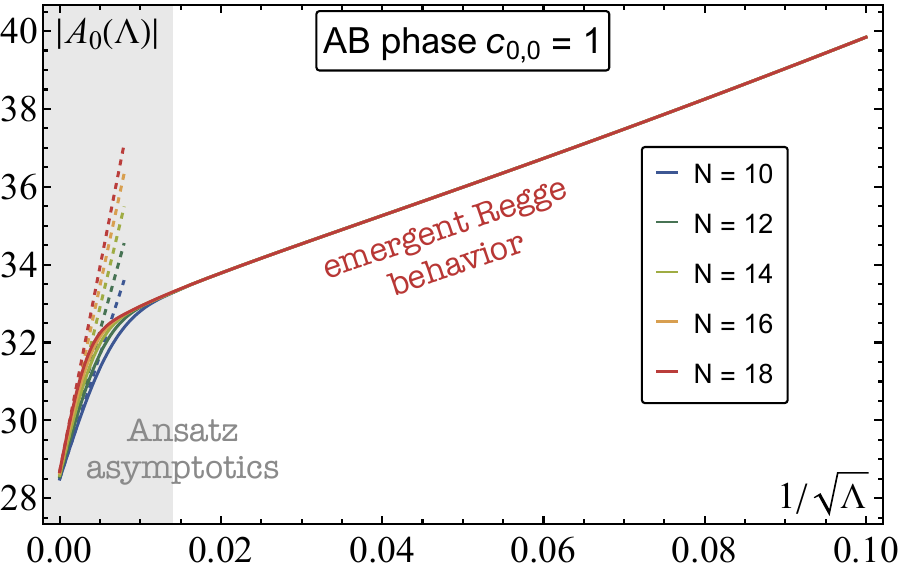}
  \hspace{1cm}
  \includegraphics[width=0.44\linewidth]{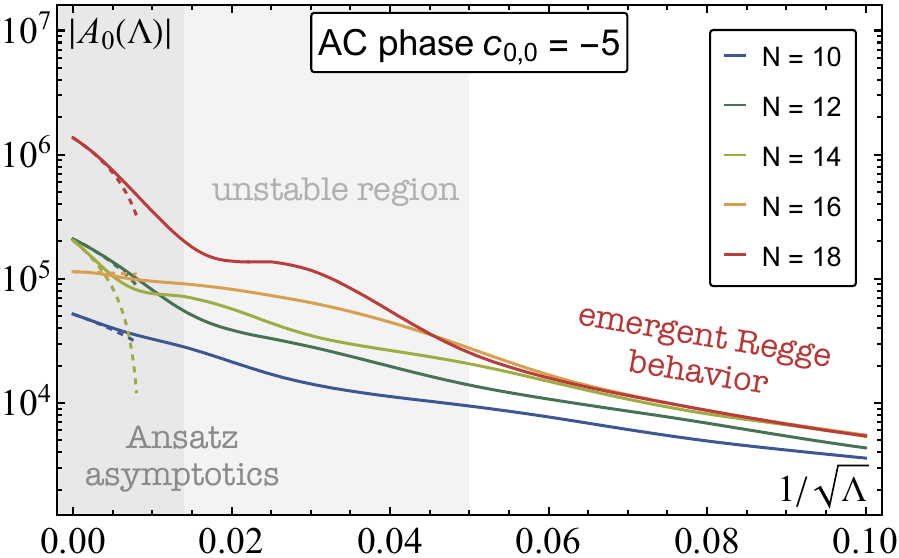}
  \caption{{\bf Left}: Regge moment $A_0(\Lambda)$ as a function of inverse energy $1/\sqrt{\Lambda}$ for an amplitude in the $AB$ phase. {\bf Right}: The same observable for an amplitude in the $AC$ phase. The dashed lines show the linear approximation to the bootstrap solution as $\Lambda\to\infty$.}
  \label{fig:regge_emergence}
\end{figure*}

Our first result is a sharp characterization of the allowed region in the $(c_{0,0},c_{1,0})$ plane, shown in Fig.~\ref{fig:dual_vs_primal}. Previous works explored this landscape using explicit \emph{primal} constructions of amplitudes \cite{Chen:2022nym,EliasMiro:2022xaa,deRham:2025vaq}, revealing a rich geometric structure in the space of S-matrix data. Here we complement this picture by establishing rigorous \emph{exclusion bounds} using the fixed-$t$ dual bootstrap approach \cite{Guerrieri:2021tak,Lopez:1975ca}, which determine the ruled-out region shown in red. We then construct high-precision primal amplitudes, extending the methods of \cite{Guerrieri:2024jkn}, that saturate the boundary of the allowed region shown in green. The boundary is obtained by maximizing linear functionals of the form $
c_{0,0}\cos\theta+c_{1,0}\sin\theta$
at fixed $\theta$ \cite{Cordova:2019lot}; varying $\theta$ scans the extremal amplitudes along the boundary. \footnote{In ref.~\cite{EliasMiro:2022xaa} these coefficients were denoted by $c_0$ and $c_2$.}

The remaining gap between the dual and primal regions reflects the different analyticity assumptions entering the two approaches. The dual analysis relies only on unitarity constraints compatible with axiomatic analyticity \cite{Martin:1965jj,Martin:1966zsy}, whereas the primal construction assumes maximal, or Landau, analyticity. Thus Fig.~\ref{fig:dual_vs_primal} separates what is rigorously excluded from what is explicitly realized under stronger analyticity assumptions.

The geometry of Fig.~\ref{fig:dual_vs_primal} contains several distinguished points. The origin $(c_{00},c_{10})=(0,0)$ corresponds to the free theory. Two cusp points, denoted by $B$ and $C$, correspond to amplitudes containing threshold bound states of spin zero and spin two, respectively. The latter has recently been termed the \emph{Froissart amplitude}, as it maximizes the average total cross section~\cite{Correia:2025uvc}.

Our second result is that the extremal amplitudes lying on the different boundary arcs of Fig.~\ref{fig:dual_vs_primal} exhibit distinct universal behavior. Although the low-energy dynamics of a gapped theory is usually sensitive to microscopic details, the boundary amplitudes organize themselves into robust regimes with shared spectra and high-energy properties. We identify three such \emph{universal phases}: an asymptotically weak phase dominated by virtual states, a resonance-dominated phase with Regge growth, and a phase exhibiting large-$N_c$-like decoupling of fixed-mass states. These phases describe different mechanisms by which ultraviolet consistency is realized from simple low-energy S-matrix data.

In the following we construct these extremal amplitudes and analyze the universal phases that emerge along the boundary of the scalar S-matrix space. A central diagnostic will be the way in which the high-energy behavior is encoded in numerical bootstrap amplitudes.

\section{UV physics and Regge moments}
\label{sec:Regge}

\noindent A basic question is how the extremal amplitudes realize their ultraviolet behavior. In a local relativistic theory one expects the high-energy, fixed-$t$ regime to be governed by Regge physics, with amplitudes behaving at high energy as $T(s,t)\sim \beta(t)s^{\alpha(t)}$. At the same time, the numerical bootstrap amplitudes are represented as truncated sums by the standard crossing-symmetric $\rho$-ansatz~\cite{Paulos:2017fhb},
\begin{equation}
T^\textbf{ans}
=\sum_{a+b+c\leq N} \alpha_{(abc)} \rho_s^a \rho_t^b \rho_u^c
\underset{\substack{s\to\infty \\ t\ \text{fixed}}}{\longrightarrow}
\beta_0(t)+\frac{\beta_1(t)}{\sqrt{s}}+\dots,
\label{gamp}
\end{equation}
where $\rho_s$ maps the cut $s$-plane to the unit disk.\footnote{See Appendix \ref{sec:numerics} for the definition of the ansatz and the conformal mapping used in the numerics.}
Numerical bootstrap solutions display clear evidence of emergent Regge trajectories~\cite{Bose:2020cod,Guerrieri:2022sod,Acanfora:2023axz, Correia:2025uvc}, indicating that the $\rho$-ansatz can approximate power-law growth over a wide range of energies. This growth should emerge at energies larger than the masses and resonance scales present in the amplitude, but not at strictly asymptotic energies: at any finite $N$, Eq.~\eqref{gamp} implies a definite large-$s$ behavior incompatible with exact Regge growth. The relevant question is therefore whether, as $N$ is increased, the window in which the amplitude displays Regge behavior becomes parametrically large before the finite-$N$ asymptotics of the ansatz takes over.

\begin{figure*}[t!]
    \centering
    \hspace{.5cm}
    \includegraphics[width=0.4\linewidth]{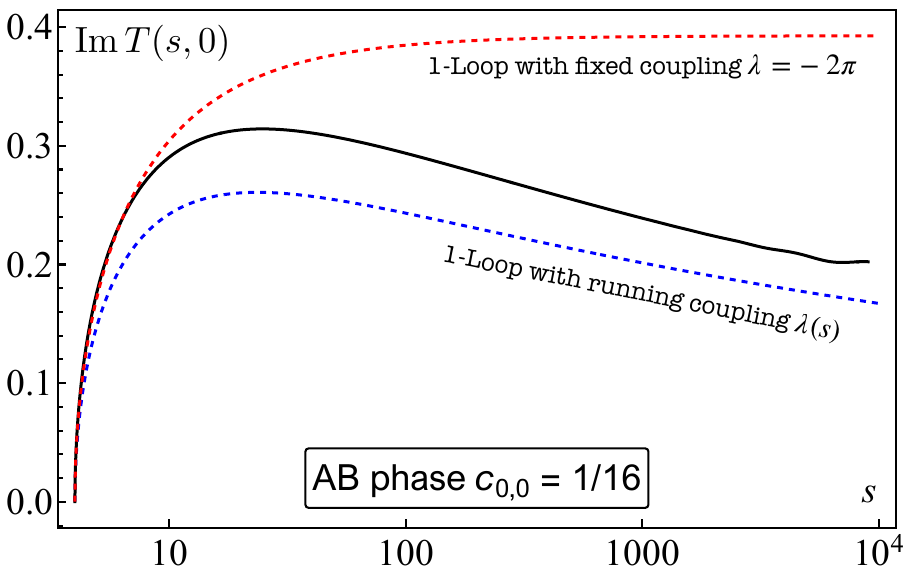}
    \hspace{1.5cm}
    \includegraphics[width=0.454\linewidth]{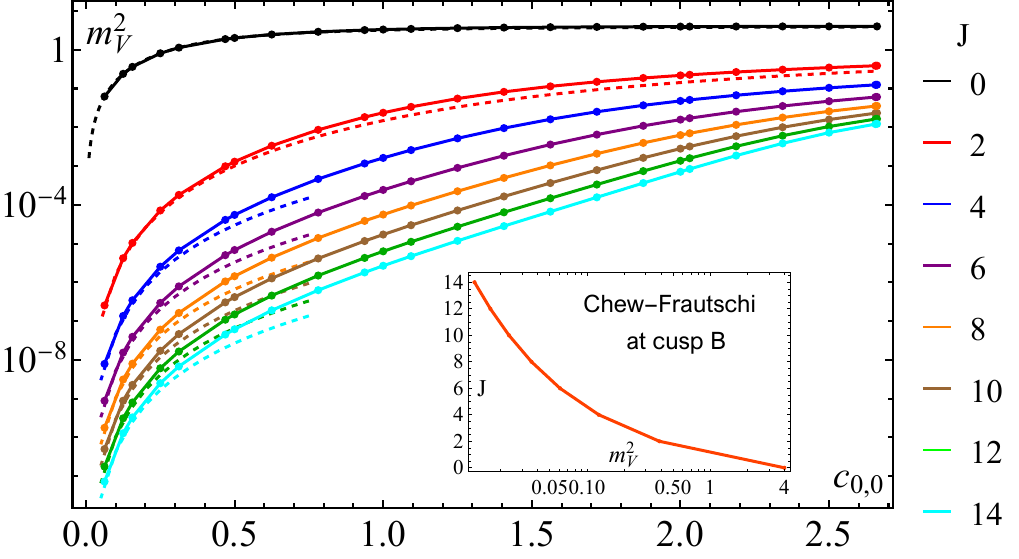}
    \caption{{\bf Left:} Black solid line: Bootstrap solution. Red dashed: one-loop $\Im T(s,0)$ from perturbation theory with $\lambda$ matched at tree-level. Blue-dashed, one-loop $\Im T(s,0)$ with running coupling \eqref{eq:running_lambda}. {\bf Right: }Positions of virtual states (a zero of the partial wave $S_\ell$ in $[0,4m^2]$) as a function of $c_{0,0}>0$. Dashed lines are perturbation theory predictions.}
    \label{fig:virtual_states}
\end{figure*}

To detect this emergent Regge behavior, we introduce \emph{Regge moments}, analogous to the Regge sum rules of Ref.~\cite{Haring:2023zwu} and complementary to the IR moments of Ref.~\cite{Bellazzini:2020cot}
\begin{equation}
A_t^n(\Lambda)=\frac{1}{2\pi i}
\oint_{|s-2+t/2|=\Lambda}
\frac{ds}{(s-2+t/2)^n}\,
T(s,t).
\label{eq:regge_moments}
\end{equation}
For $n=1$, we suppress the label, and it becomes
\begin{equation}
A_t(\Lambda)=
\frac{1}{\pi}
\int_0^\pi d\theta\,
\Re\,T\!\left(2-\frac{t}{2}+\Lambda e^{i\theta},t\right),
\end{equation}
and if $T(s,t)\sim \beta(t)s^{\alpha(t)}$, the large-$\Lambda$ scaling determines an effective Regge exponent
\begin{equation}
\alpha^\text{eff}(t)
\sim
\Lambda\frac{\partial}{\partial\Lambda}\log | A_t(\Lambda)| .
\label{eq:alpha_eff}
\end{equation}

Fig.~\ref{fig:regge_emergence} shows $A_0(\Lambda)$ as a function of inverse energy $1/\sqrt{\Lambda}$ for increasing truncation order $N$. The left panel corresponds to an amplitude on the $AB$ arc; other amplitudes on this arc exhibit qualitatively similar behavior. A broad intermediate region displays stable scaling before the intrinsic large-$\Lambda$ asymptotics of the finite-$N$ $\rho$-ansatz takes over. This ultimate behavior is fixed by the coefficients $\beta_0(t)$ and $\beta_1(t)$ in Eq.~\eqref{gamp}, and controls the curves only extremely close to $1/\sqrt{\Lambda}=0$. As $N$ increases, the scaling region expands, signalling convergence toward a genuine non trivial Regge behaviour. In this example the effective intercept approaches zero smoothly, $\alpha^\text{eff}(0)\to 0$, with no evidence for a negative asymptotic plateau. 


The right panel shows a more extreme example, representative of the positive-intercept arcs $AC$ and $BC$, where convergence is slower. $A_0(\Lambda)$ displays a clear tendency to grow with energy. At intermediate energies the Regge regime progressively stabilizes as $N$ increases, while at higher energies an unstable non-monotonic region remains visible and convergence in $N$ has not yet been reached. Only at extremely large $\Lambda$ does the exact asymptotic behavior of the $\rho$-ansatz finally dominate. In the stable Regge regime, the effective intercept develops a clear plateau around $\alpha^\text{eff}(0)\approx 1$. Similar behavior is observed throughout the $BC$ and $AC$ arcs.\footnote{In particular, we find that $c_{0,0}<0$ is always associated with $\alpha_\text{eff}(0)>0$, consistent with Fig.~5 of Ref.~\cite{EliasMiro:2022xaa} and Fig.~10 of Ref.~\cite{deRham:2025vaq}.} Further details are given in Appendix~\ref{app:regge}.




\section{Universal phases}

\noindent The Regge moments reveal that the boundary of Fig.~\ref{fig:dual_vs_primal} is not a featureless curve of extremal amplitudes. Instead, it splits into three arcs with distinct ultraviolet behavior. Along $AB$, amplitudes exhibit marginal Regge scaling, with intercept approaching zero. Along $AC$ and $BC$, which meet at the Froissart cusp $C$, amplitudes display genuine Regge growth with positive intercept.

This ultraviolet organization is accompanied by a corresponding organization of the spectrum. We now show that amplitudes on each arc share universal spectral features, involving virtual states, resonances, and large-$N_c$-like decoupling. This motivates the interpretation of the three boundary arcs as distinct universal phases of four-dimensional scalar scattering.\footnote{Additional details on the extraction of virtual states, resonance poles, and threshold diagnostics are given in Appendix~\ref{app:numanal}.}

\subsection*{AB phase: asymptotically free scalar theories}

\noindent We begin with the amplitudes populating the boundary arc $AB$, which smoothly interpolates between the free theory and B cusp. A striking feature of these amplitudes is the absence of resonant peaks in the cross-section, see left plot of  Fig.~\ref{fig:virtual_states}. Instead, these amplitudes exhibit an infinite tower of \emph{virtual states}.

\begin{figure*}
\centering
\includegraphics[width=\textwidth]{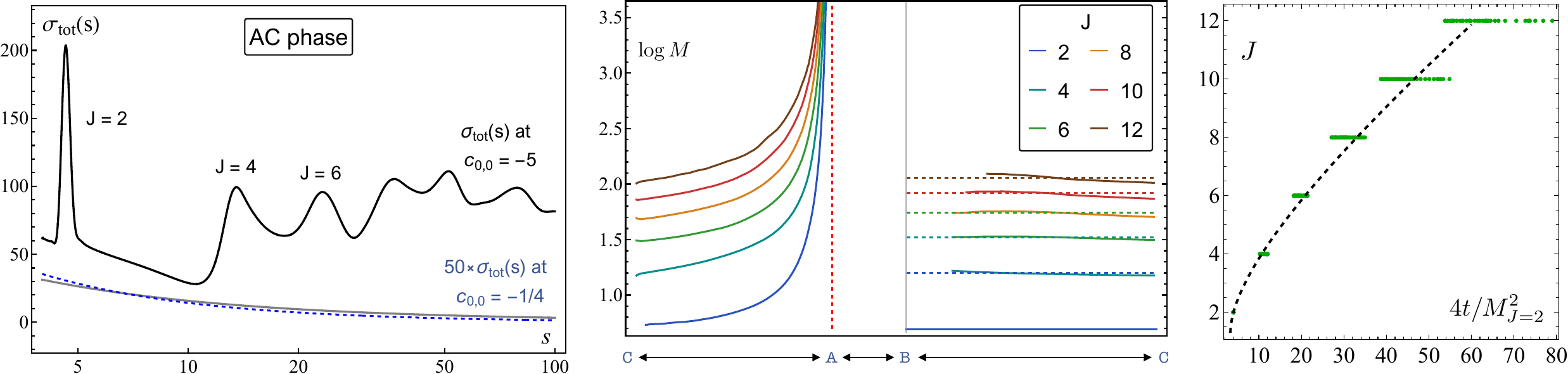}
\caption{
Energy decoupling of the higher-spin sector along the $AC$ arc.
{\bf Left}: Total cross-section at two points on the boundary. Near the $C$ cusp, $c_{0,0}=-5$, the amplitude displays a rich pattern of higher-spin resonances, while near the free point, $c_{0,0}=-0.25$, the low-energy cross-section follows the perturbative EFT prediction shown by the blue dashed curve.
{\bf Middle}: Pole masses on the leading trajectory, extracted from partial-wave zeros using $s=(M+i\Gamma/2)^2$, along the boundary arcs. Approaching the free point $A$ from the $AC$ arc, the higher-spin spectrum is pushed to high energy, whereas along $BC$ the masses remain finite.
{\bf Right}: The leading trajectory after rescaling by the spin-two mass (green dots). Its approximate stability shows that the $AC$ arc realizes a collective, self-similar energy decoupling of a strongly interacting Regge sector.
}
\label{fig:AC_cross_sections}
\end{figure*}

A virtual state is   a zero of the partial-wave $S_\ell(s)$ for $s\in[0,4m^2]$. This zero is in correspondence with a pole on the second Riemann sheet below the two-particle threshold.~\footnote{We remind the reader that $S_\ell=1+i\tfrac{\sqrt{s-4}}{\sqrt{s}}f_\ell(s)$, and $f_\ell(s)=\tfrac{1}{32\pi}\int_{-1}^1 dx P_\ell(x)T(s,\tfrac{4-s}{2}(1-x))$. See also Appendix \ref{sec:numerics}.} Such states are not exotic: they arise already in simple quantum mechanical potentials~\cite{Newton1982}, and a well-known example in nuclear physics is the $^1 S_0$ virtual state in $np$ scattering, with binding energy $\sim 60\,\text{keV}$~\cite{Weinberg:1990rz}.

The origin of these states can be understood directly from the low-energy scalar EFT. At tree level the amplitude is constant,
$
T(s,t,u)=32\pi c_{0,0}
$,
which yields for the spin-0 partial wave
$
S_0 = 1-2\,\frac{\sqrt{4-s}}{\sqrt{s}}\,c_{0,0}
$.
Setting $S_0=0$ predicts a virtual-state of mass
\begin{equation}
m_V^2=\frac{16c_{0,0}^2}{1+4c_{0,0}^2} \, ,
\end{equation}
in excellent agreement with numerics, see right plot of Fig.~\ref{fig:virtual_states}. As $c_{0,0}\to 0$, the virtual state becomes massless, while at strong coupling its mass approaches the two-particle threshold. At the B cusp, the spin zero virtual state becomes a stable threshold bound state.

Higher-spin virtual states arise from loop corrections (see Appendix~\ref{app:one-loop}). We find quantitative agreement with perturbation theory for spin two, while deviations increase for higher spins, signalling the onset of genuinely nonperturbative physics controlling the tail of higher-spin partial waves.

The weakening of interactions at high energy can be estimated from the one-loop running coupling,
\begin{equation}
\lambda(s)
=
-\frac{32\pi c_{0,0}}
{1+\frac{3}{\pi}c_{0,0}\log(3s/4)}
\sim
-\frac{32\pi^2}{3\log(3s/4)},
\label{eq:running_lambda}
\end{equation}
where we matched at the crossing-symmetric point.
This logarithmic behaviour suggests an asymptotically weak regime.

The full amplitude exhibits a scaling behaviour compatible with perturbation theory. Perturbatively, both tree-level and loop corrections decay as $\log^{-1}(s)$, and so we expect in general at any loop order. Numerically, we observe both in the Regge and fixed angle limit a slow log decay of the amplitude -- see Appendix~\ref{appABphase}.

This behaviour is consistent across the entire arc $AB$ and with the Regge moments analysis, indicating a universal UV regime characterised by:  logarithmically weak interactions, the absence of resonances,  and the presence of an infinite tower of virtual states.
 
\subsection*{AC phase: Strongly coupled higher-spin sector}

\noindent Moving to the left of the free theory, we enter a region of growing amplitudes whose low-energy description is a scalar EFT with positive quartic coupling. The scalar interaction is then repulsive and does not produce a spin-zero virtual state. Higher-spin channels, however, remain nontrivial. At one loop the virtual-state condition takes the form
\begin{equation}
S_{\ell\geq 2}(s)
=
1-
\frac{\sqrt{s-4}}{\sqrt{s}}\,
\lambda^2 \mathcal{I}^{1}_\ell(s)=0 ,
\end{equation}
where $\mathcal{I}_\ell^1$ is the partial-wave projection of the one-loop amplitude.
Since the condition depends only on $\lambda^2$, it is insensitive to the sign of the quartic coupling. Thus the spin-zero virtual state disappears on the repulsive side, while higher-spin virtual states persist.

 \begin{figure*}
    \centering
    \includegraphics[width=0.422\linewidth]{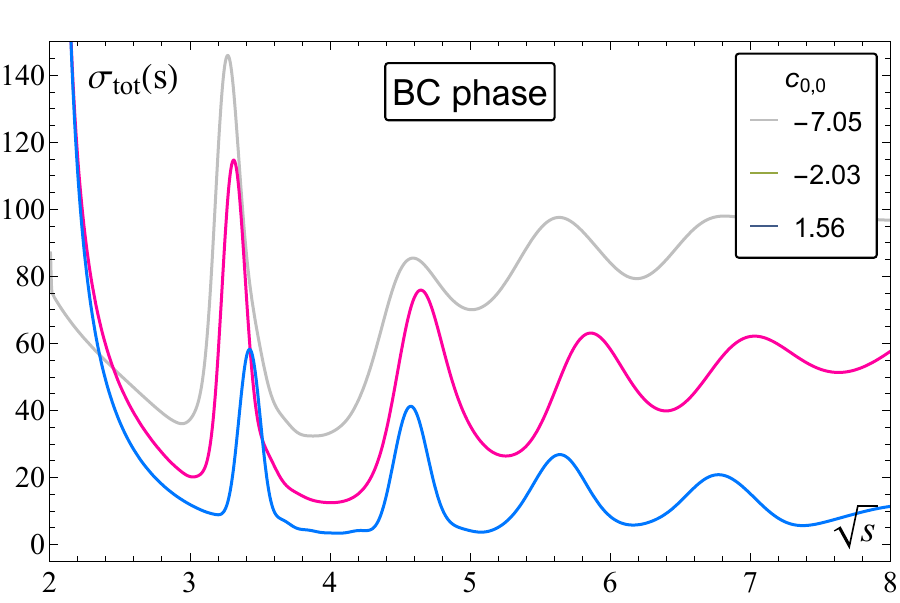}
\hspace{0.cm}
\includegraphics[width=0.275\linewidth]{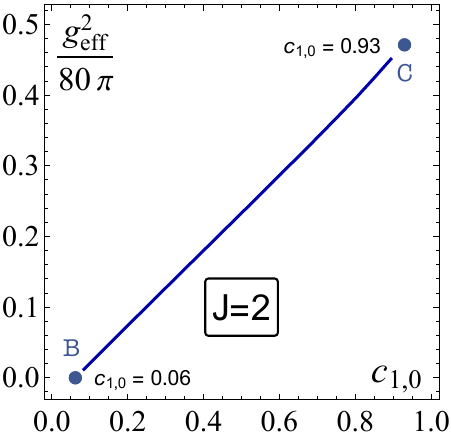}
    \includegraphics[width=0.27\linewidth]{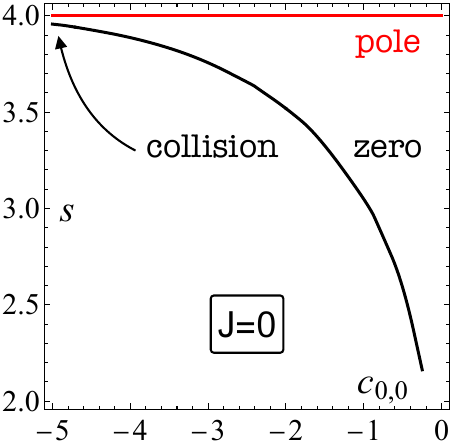}
    \caption{{\bf Left:} Cross section on the upper branch. {\bf Center \& Right:} The two mechanisms that decouple the  threshold poles at the cusps that surround the $BC$ region. Center: effective coupling of the spin two threshold pole as function of $c_{1,0}$. Right: the mass of the virtual state as function of $c_{0,0}$ showing how it approaches the threshold pole.}
    \label{fig:bcfigs}
\end{figure*}

Amplitudes along the $AC$ arc exhibit a rich spectrum of massive resonances. 
The endpoint $C$ is the Froissart amplitude, whose spectrum was analyzed in detail in Ref.~\cite{Correia:2025uvc}. 
Tracking the poles along the full arc reveals a simple mechanism: the higher-spin sector remains strongly interacting, but its characteristic energy scale is pushed upward as one approaches the free point $A$. 
This energy decoupling is directly visible in the total cross-section, shown in the left panel of Fig.~\ref{fig:AC_cross_sections}. 
Near the $C$ cusp the cross-section displays a dense pattern of peaks and valleys, while moving toward $A$ shifts these structures to higher energies and leaves a progressively smoother low-energy cross-section, approaching the $+\lambda\phi^4$ EFT expectation.

The lightest spin-two resonance on the leading Regge trajectory provides a useful measure of this scale separation. 
In the middle panel of Fig.~\ref{fig:AC_cross_sections} we show pole positions extracted from zeros of the partial waves, writing
$s=(M_\ell+i\Gamma_\ell/2)^2$. 
Moving from the $C$ cusp toward the free point $A$, the leading spin-two pole is rapidly pushed to high energy. 
Numerically, its mass is consistent with a steep power-law growth, approximately $M_2\sim |c_{1,0}|^{-3/2}$ close to the free point, while $c_{1,0}\sim c_{0,0}^2$ as predicted by perturbation theory. 
Its width grows with the same characteristic scale, with no evidence that $\Gamma_2/M_2$ tends to zero.

This behavior is not special to the lightest spin-two state. 
Higher-spin resonances, including states on subleading Regge trajectories, are shifted to high energy in the same way. 
Moreover, after normalizing the spectrum by the spin-two scale $M_2$, the leading trajectory is approximately unchanged, as shown in the right panel of Fig.~\ref{fig:AC_cross_sections}. 
Thus the higher-spin sector decouples in an essentially self-similar way: its overall energy scale is sent to infinity, while its internal Regge structure and strong-coupling character persist. 
The $AC$ arc therefore realizes energy decoupling of a strongly interacting Regge sector, rather than weak-coupling decoupling through narrow resonances.

\subsection*{BC phase: threshold bound states and large-$N_c$-like decoupling}

\noindent The $BC$ arc is special because the spin-zero and spin-two threshold poles coexist in its interior, while at the endpoints only one of them remains. The cusp $B$ contains the scalar threshold state, whereas the cusp $C$ contains the spin-two threshold state. Thus the arc interpolates between two threshold theories through two inequivalent mechanisms: the scalar pole is removed near $C$ by a zero-pole cancellation, while the spin-two and higher-spin sector decouples near $B$ by a large-$N_c$-like suppression of residues at approximately fixed mass.

Let us first describe the scalar mechanism. Starting from the $B$ cusp, the spin-zero bound state can be followed along the $BC$ arc. At $c_{0,0}=0$, a virtual-state zero emerges from the subthreshold region and moves toward the two-particle threshold. As the $C$ cusp is approached, this zero collides with the scalar threshold pole and cancels its effect, rightmost panel in Fig. \ref{fig:bcfigs}. This provides a natural continuation to the $AC$ arc, where the repulsive scalar interaction no longer supports a spin-zero virtual state. A simple CDD model of this zero-pole collision is given in Appendix~\ref{app:cdd}.

The higher-spin sector behaves differently. Rather than being pushed to high energies, as on the $AC$ arc, the resonance masses remain approximately fixed along $BC$ -- see middle panel of Fig. \ref{fig:AC_cross_sections}, while their widths and residues decrease.  This visible in the left panel of Fig.~\ref{fig:bcfigs}: the peaks in the total cross-section, corresponding to resonances of increasing spin, stay at nearly fixed positions as $c_{0,0}$ is varied, but become progressively narrower. For example, the first sharp peak is associated with a spin-four resonance. This pattern suggests a large-$N_c$-like decoupling mechanism. As in large-$N_c$ gauge theories, the spectrum remains at finite mass while the couplings of the resonances to the low-energy degrees of freedom are suppressed.

The threshold bound states provide a useful quantitative probe of this large-$N_c$-like decoupling. Using the $K$-matrix analysis of Appendix~\ref{sec:K-matrix}, a spin-$\ell$ threshold bound state modifies the partial wave as
\begin{equation}
f_\ell(s)
\sim
- g_{\rm eff}^2 (s-4)^{\ell-1},
\label{anom}
\end{equation}
instead of the standard scattering-length behavior
\begin{equation}
f_\ell(s)
\sim
a_\ell (s-4)^\ell,
\qquad a_\ell \ge 0 \quad (\ell\ge2).
\end{equation}
The anomalous sign and modified threshold scaling in Eq.~\eqref{anom} define an effective threshold coupling $g_{\rm eff}$. In the center plot of Fig.~\ref{fig:bcfigs}, we show that the effective coupling of the spin-two threshold state vanishes linearly as $c_{1,0}$ decreases. A similar suppression is observed for higher-spin resonances, although extracting their residues is numerically harder because the corresponding poles lie at complex masses.

The linear relation between $c_{1,0}$ and the effective coupling has a simple dispersive interpretation. If the $c_{1,0}$ sum rule is dominated by a spin-two pole of mass $M_2$ and coupling $g$, then
\begin{equation}
    c_{1,0}
    =
    \frac{1}{\pi}\int_4^\infty ds^\prime\,
    \frac{\Im T(s^\prime,4/3)}{(s^\prime-4/3)^3}
    \sim
    \frac{g^2}{M_2^6}.
\label{eq:c10_disp}
\end{equation}
Along the $BC$ arc, $M_2$ remains approximately fixed, so $c_{1,0}$ measures the residue rather than a moving mass scale. This is precisely the large-$N_c$-like pattern: states remain fixed in the spectrum, while their couplings to the low-energy scalar degrees of freedom are suppressed. The endpoint value is reproduced by a threshold model with leading imaginary part $4/\sqrt{s-4}$, which gives
$c_{1,0}=27\sqrt{3/2}/512\simeq0.06$.

The $BC$ arc therefore realizes a different kind of Regge phase. The higher-spin spectrum remains at approximately fixed mass, but its residues decrease toward $B$, producing a large-$N_c$-like weakening of the interactions. The scalar threshold state instead disappears near $C$ through a zero-pole cancellation. In this way, the arc connects the scalar and spin-two threshold cusps without pushing the Regge spectrum to high energy.

This completes the classification of the three universal regimes on the boundary: virtual-state accumulation on $AB$, energy decoupling on $AC$, and large-$N_c$-like decoupling at fixed mass on $BC$.

\section{Outlook}

\noindent \noindent A natural question raised by this work is what microscopic theories, if any, realize the extremal amplitudes that populate the boundary of the island.
The universal phases identified here exhibit qualitatively distinct mechanisms for the UV completion of scalar EFTs: asymptotically weak amplitudes with virtual states, resonance-dominated spectra with Regge growth, and large-$N_c$-like decoupling in which interactions vanish at fixed mass. Understanding whether these mechanisms can emerge from microscopic Lagrangian theories remains an open problem.

A suggestive possibility is provided by confining gauge theories with a light pseudoscalar mode analogous to the $\eta'$ -- e.g.   $SU(N_c)$ gauge group with $N_f=1$ flavor.  In such theories, the anomalous axial symmetry and large-$N_c$ dynamics generate a nontrivial effective potential \cite{Veneziano:1979ec, Witten:1980sp, DiVecchia:1980yfw}, while the UV spectrum contains towers of higher-spin resonances. It would be interesting to understand whether some regions of the island, or perhaps regions reached by scanning over higher-order S-matrix data, admit an interpretation of this kind. In Appendix~\ref{sec:3dshape} we give a preliminary map of the $(c_{0,0},c_{1,0},c_{0,1})$ space, which suggests that a candidate for large-$N_c$ $\eta'$ dynamics may lie on a higher-dimensional boundary of the space of scalar amplitudes.

Another possibility is that some of these amplitudes arise in higher-dimensional constructions, for instance as effective theories of Goldstone modes localized on branes in presence of a potential that gaps them, where additional states provide the required UV completion \cite{Leigh:1989jq,Nicolis:2008in,Cheung:2015ota}. Alternatively, the bootstrap may reveal solutions of the QFT constraints that do not admit a conventional weakly coupled Lagrangian description.

A limitation of the present analysis is that the amplitudes considered here do not include explicit particle-production channels. The absence of production is compatible with the elastic constraints imposed here, but it may fail to survive the full bootstrap problem once multiparticle channels are included. It would therefore be important to extend this study using approaches that incorporate inelastic processes, such as those developed in \cite{Tourkine:2021fqh,Tourkine:2023xtu,Gumus:2024lmj}. This could help determine which regions of the island correspond to physically realizable theories and which instead reflect limitations of elastic bootstrap ansätze.

Charting higher-dimensional projections of the island, extending the analysis to systems with continuous global symmetries, fermions, or gauge bosons, and classifying the associated phases are natural next steps toward understanding the space of consistent QFTs.

 \section*{Acknowledgements}
 We thank Miguel Correia, Alessandro Georgoudis, Kelian Haring, Zohar Komargodski, Leonardo Rastelli, Pedro Vieira, Piotr Tourkine, Alexander Zhiboedov for useful discussions.
 JEM  is supported by the European Research Council, grant agreement n. 101039756.
 A.G. is supported by a Royal Society University Research Fellowship, URF/R1/241371.



\small

\bibliographystyle{utphys}

\bibliography{biblio}



\clearpage
\appendix
\onecolumngrid

\section{Conventions and numerical setup}
\label{sec:numerics}

In this appendix, we explain our conventions and provide a full account of our numerical bootstrap setup. 

We define partial wave coefficients $f_\ell$ of the amplitude via
\be
T(s,t) = \sum_{\ell=0}^\infty n_\ell \, f_\ell(s) \, P_\ell (1{+}\tfrac{2t}{s-4}) \quad \text{with} \quad
f_\ell(s) = \frac{1}{32\pi}\int^1_{-1} dz \, P_\ell(z) M(s,t(z))
\label{eq:pwave_exp}
\ee
where $P_\ell$ are Legendre polynomials, $n_\ell=(16\pi)(2\ell+1)$, and $z$ is the scattering angle which satisfies $t(z)=(-1/2)(s-4)(1-z)$ and $u(z)=t(-z)=(-1/2)(s-4)(1+z)$. 
Odd-$\ell$ partial wave coefficients vanish due to $t \leftrightarrow u$ symmetry of the amplitude.

\subsection{Primal bootstrap}
\label{app:primal}
We use the crossing-symmetric ansatz for the amplitude
\be
\label{eq:ansatz}
\begin{aligned}
T^\text{ans}(s,t) = \, &\alpha_\text{th} \left( \frac{1}{\rho_{20/3}(s)-1} + \frac{1}{\rho_{20/3}(t)-1} + \frac{1}{\rho_{20/3}(u)-1} \right) 
+ \sum_{\sigma \in \Sigma} \,\, \sum_{a \leq N_\sigma} \alpha_a \left( \rho_\sigma(s)^a + \rho_\sigma(u)^a + \rho_\sigma(t)^a \right)  \\
&+ \sum_{\sigma \in \Sigma} \,\, \sum_{a+b \leq N_\sigma} \,\, \alpha_{(a,b)} \left(
    \rho_\sigma(s)^a \rho_\sigma(t)^b + \rho_\sigma(t)^a \rho_\sigma(u)^b + \rho_\sigma(u)^a \rho_\sigma(s)^b \ 
\right) \ ,
\end{aligned}
\ee
where the wavelet term $\rho_\sigma(s)$ centered at $\sigma$ is given by
\be
\rho_\sigma(s) = \frac{\sqrt{\sigma-4}-\sqrt{4-s}}{\sqrt{\sigma-4}+\sqrt{4-s}} \quad , \quad \sigma>4 .
\ee
Note that a wavelet function goes to a constant $\rho_\sigma(s) \to -1$ at large energies $|s| \to \infty$, while $\rho_\sigma(8{-}\sigma)=0$, $\rho_\sigma(4)=1$ and $\rho_\sigma(\sigma)=-i$. Physical energies in its argument map to the boundary of unit disk, $\rho_\sigma : (4+ i \epsilon,\infty+i\epsilon) \mapsto \{ e^{i\theta} \, | \, 0 < \theta < \pi \}$.

The first term of~\eqref{eq:ansatz} corresponds to a possible singular behavior near the two-particle production threshold $\lim_{s\to4} T(s,t) \sim \alpha_\text{th} \, \frac{i \sqrt{2/3}}{\sqrt{s-4}} + O(s-4)$, and its coefficient is bound by unitarity to take two values $\alpha_\text{th} \in \{0,64\pi/\sqrt{2/3}\}$. For the regular terms of~\eqref{eq:ansatz}, we use the following centers and powers in the outer sums:
\be
\Sigma = \{ 20/3,10,20,30,40,50,60,86 \} \quad , \quad N_\sigma= \begin{cases}
N_\text{max} \qquad \quad \text{if} \quad \sigma=20/3, \\
N_\text{max}{-}2 \qquad \quad \text{otherwise}.
\end{cases}
\ee
In primal approach, we solve the semidefinite programming problem for the unknowns $\{\alpha_\text{th}, \alpha_a, \alpha_{a,b}\}$, of finding extremal values of the Wilson coefficients $c_{n,m} \sim \partial^n_\sigma \partial^m_\tau T(s,t)$ with respect to partial wave unitarity, thereby constructing the function $T^\text{ans}(s,t)$. Unitarity condition is imposed on a finite number of partial wave coefficients
\be
\label{eq:pw_unitarity}
\bigg| 1 +i \frac{\sqrt{s-4}}{\sqrt{s}} f_\ell(s) \bigg|^2 \leq 1 \quad \text{for} \quad \ell\leq L_\text{max} .
\ee
We sample \reef{eq:pw_unitarity}
on a number of grid points $s_\text{max}$, and recast it as a $2\times2$ matrix
\be
    \mathcal{U}_\ell =
    \begin{pmatrix}
        1+\rho \, \text{Re}f_\ell & 1-\rho \,\text{Im}f_\ell \\
        1-\rho \,\text{Im}f_\ell & 1-\rho \,\text{Re}f_\ell
    \end{pmatrix}
    \succeq 0 \, ,
    \label{eq:unitarity_2x2}
\ee
which is linear in the primal variables $\{\alpha_\text{th}, \alpha_a, \alpha_{a,b}\}$, and $\rho^2(s)=\sqrt{(s-4)}/\sqrt{s}$ is the two-body phase space factor.
We feed the semi-definite unitarity constraints~\eqref{eq:unitarity_2x2} together with the linear constraint $-c_{0,0}\sin\theta + c_{1,0}\cos\theta = \text{const.}$ to SDPB~\cite{Simmons-Duffin:2015qma,Landry:2019qug} to obtain the solutions extremizing $c_{0,0}\cos\theta + c_{1,0}\sin\theta$. Analogous linear combinations of Wilson coefficients are used when probing the orthogonal directions, as we did in Appendix~\ref{sec:3dshape}. In the entirety of our numerical runs, we used the parameters $N_\text{max}=\{10,12,14,16,18,20\}$, $L_\text{max} \in \{16,18\}$ and $s_\text{max}=300$ to ensure convergence.

To improve the convergence along higher-spin dominated phases, we also implement the \emph{subtracted positivity constraints}~\cite{Guerrieri:2021ivu,EliasMiro:2022xaa}
\be
\text{Im}T(s,t) - \sum_{\ell=0}^{L_\text{max}}(16\pi)(2\ell+1)\text{Im}f_\ell(s) \geq 0 \, .
\ee
on the $s$-grid for $t \in \{0.27, 1, 2, 3, 3.73, 3.99994, 3.99996, \
3.99998, 3.99999, 4.00000\}$.

\newpage
\subsection{Dual bootstrap}

Here we describe how to construct the dual for the optimization problem described in App.~\ref{app:primal} by writing a suitable Lagrangian implementing analyticity, unitarity and crossing symmetry constraints. See also~\cite{Guerrieri:2021tak} for further details.

\paragraph{Analyticity.} We start by writing the double-subtracted, fixed-$t$ dispersion relation for the amplitude
\be
T(s,t) = T(s_0,t_0) + \int^\infty_{4} \!\!\!\! dv \, K(v,s,t;t_0) \, \text{Im} T(v,t) + \int^\infty_{4} \!\!\!\! dv \, K(v,t,t_0;s_0) \, \text{Im} T(v,t_0)
\label{eq:fixedt}
\ee
where the kernel is given by $
K = \frac{1}{\pi} \left( \frac{1}{v-s} + \frac{1}{v-4+s+t} - \frac{1}{v-t_0} - \frac{1}{v-4+t+t_0} \right)$.
This relation is valid for $-28 < t < 4$ and any $s \in \mathbb{C}$. After choosing $s_0=t_0=4/3$ so that the subtraction constant becomes $c_{0,0}=T(4/3,4/3)$ and projecting \eqref{eq:fixedt} on the $\ell$-th partial wave at a fixed value of $s$, we obtain the \emph{Roy equations}
\be
a_j(s) = \text{Re} f_j(s) - 2 \, c_{0,0} \, \delta_{0,j} - \text{P.V.} \int^\infty_4 dv \sum^\infty_{\ell = 0} k_{j,\ell}(s,v) \, n_\ell \, \text{Im}f_\ell(v) 
= 0 \, ,
\label{const_analyticity_o1}
\ee
where the projected kernel $k_{j,\ell}$ is given by
\be
k_{j,\ell}(s,v) = \frac{1}{32 \pi}  \int^1_ {-1} dz\,  P_j(z) \, \Big[ P_\ell(1{+}\tfrac{2t(z)}{v-4}) K(v,s,t(z);4/3) + P_\ell(1{+}\tfrac{8/3}{v-4}) K(v,t(z),4/3;4/3) \Big] \, .
\label{fixedt_partial_kernel_o1}
\ee
Notice that the kernel has a singularity $k_{j,\ell}(s,v) \sim \frac{\delta_{j,\ell}}{n_\ell} \, \frac{1}{v-s}$. Convergence conditions for the partial wave expansion of the amplitude and feasibility for the dual problem require a limited range of validity for Roy equations in the $s$ variable, $4 \leq s \leq \mu^2$ for $\mu^2=12$. See~\cite{EliasMiro:2025rqo} for an alternative approach to derive Roy-like equations with $\mu^2 \to \infty$ via crossing-symmetric dispersion relations instead of fixed-$t$.
\paragraph{Crossing symmetry.}
Fixed-$t$ dispersion relation \eqref{eq:fixedt} is manifestly $s \leftrightarrow u$ symmetric, but it is not the case for $s \leftrightarrow t$. A crossing equation $\mathcal{F}(s,t) = M(s,t) - M(s,4{-}s{-}t)=0$ is needed to impose full crossing symmetry. Plugging \eqref{eq:fixedt} in the crossing equation and projecting on partial waves give
\bea
&\mathcal{F}(s,t) = \int_{4}^\infty dv \sum_{\ell=0}^\infty  F_\ell(v,s,t;t_0,s_0) \, n_\ell \, \text{Im} f_\ell(v) = 0 \, \label{crossing_full_o1}
\eea
with the kernel $F_\ell$ given by 
$
F_\ell = P_\ell(1{+}\tfrac{2t}{v-4})K(v,s,t;t_0){-}P_\ell(1{+}\tfrac{2u}{v-4})K(v,s,u;t_0)+ P_\ell(1{+}\tfrac{2t_0}{v-4})(K(v,t,t_0;s_0){-}K(v,u,t_0;s_0))$.
Notice that $F_{\ell=0}$ vanishes, meaning that the crossing nontrivially relates spins $\ell \geq 2$ to each other. We will take derivatives of \reef{crossing_full_o1} around $\bar{s}=\bar{t}=\bar{u}=0$ to introduce the crossing constraints in a systematic manner to the optimization problem.
\bea
\mathcal{F}^\textsf{(n,m)} &= \int_{4}^\infty dv \sum_{\ell \text{ even}}^\infty F^\textsf{(n,m)}_\ell(v) \, n_\ell \, \text{Im} f_\ell(v) = 0 \quad \text{where} \quad F^\textsf{(n,m)}_\ell(v) = \frac{\partial }{\partial t^n}\frac{\partial }{\partial s^m}F_\ell(v,s,t;\tfrac{4}{3},\tfrac{4}{3})\big |_{s=t=4/3} \, .
\label{crossing_constraints_o1}
\eea
There are no constraints for $\textsf{n}+\textsf{m} \leq 3$, and first nontrivial sum rule appears at $\textsf{(n,m)}=(1,3)$.
\paragraph{Primal problem.} We can recast the primal problem in App.~\ref{app:primal} in terms of partial waves as the primal unknown variables via the following Lagrangian
\bea
\mathcal{L} &= c_{0,0}\cos\theta + c_{1,0}\sin\theta + \kappa_0 (k - c_{0,0}\sin\theta + c_{1,0}\cos\theta) + \kappa_2 \left[ c_{1,0} - \sum^\infty_{\ell=0} \int^\infty_{4} \!\!\!\! dv \frac{1}{\pi} \frac{n_\ell \, \text{Im} f_\ell(v)}{(v-4/3)^3} \right] \label{disp_c2} \\
&+ \sum^{\textsf{J}}_{j=0} \int^{\mu^2}_4 \!\!\!\! ds \, w_j(s) \, a_j(s)
+ \sum_{\textsf{n,m}} \nu_\textsf{n,m} \, \mathcal{F}^\textsf{(n,m)}
+ \sum^\infty_{\ell=0} \int_4^\infty \!\!\!\! dv \, n^2_\ell \, \lambda_\ell(v) \, \mathcal{U}_\ell(v)
\nonumber
\label{const_unitarity_o1}
\eea
where $\{\theta, k\}$ are constants and $\{\kappa_0, \kappa_2, w_j(v), \nu_\textsf{n,m}, \lambda_\ell(v)\}$ are called dual variables. 
\paragraph{Dual problem.} Integrating out primal variables leaves us with the dual Lagrangian
\bea
\mathcal{D}[\{w_j\}] = \kappa_0 k + \sum^{\textsf{J}}_{\ell=0} \int^{\mu^2}_4 \!\!\!\! \frac{dv \, n_\ell}{\rho^2(v)} \left[ \, \bar{\mu}_\ell(v) + \sqrt{ \bar{\mu}_\ell(v)^2 + ( w_\ell(v)/n_\ell )^2 } \, \right] + \sum^\infty_{\ell=\textsf{J}+2} \int^{\mu^2}_4 \!\!\!\! \frac{dv \, n_\ell}{\rho^2(v)} \left[ \, \bar{\mu}_\ell(v) + \sqrt{ \bar{\mu}_\ell(v)^2 } \, \right]
\eea
where \bea
\overline{\mu}_\ell(v) = \frac{\kappa_2/\pi}{(v-4/3)^3} + \sum_{\textsf{n,m}} \nu_\textsf{n,m} F^\textsf{(n,m)}_\ell(v) - {\overline w}_{\ell}(v) \quad \text{where} \quad {\overline w}_{\ell}(v) = \sum^{\textsf{J}}_{j=0} \text{P.V.} \int_4^{\mu^2} \!\!\!\! ds \, w_j(s) k_{j,\ell}(s,v) \, . \label{wbar_ints_o1}
\eea

The functional $\mathcal{D}[\{w_j\}]$ is convex and bounded from below, satisfying the dual property $\mathcal{D}[\{w_j\}] \geq \mathcal{L}$. We numerically find its minimum to obtain the exclusion bounds.

\newpage

\section{Emergent asymptotic Regge limit}
\label{app:regge}

The success of Regge theory can be traced back to the resolution of an apparent tension between the presence of higher-spin resonances and the Froissart bound on the high-energy behavior of scattering amplitudes. The $t$-channel exchange of a spin-$J$ particle contributes schematically as
$\tfrac{P_J((s-u)/(s+u))}{t-m_J^2}\sim s^J/t$ when $s\to\infty$ at fixed $t$. For $J>1$, this growth is incompatible with the Froissart bound $|T|\lesssim s\log^2 s$ in the forward limit $t=0$. Regge theory resolves this tension by replacing fixed-spin growth with a $t$-dependent asymptotic behavior,
\begin{equation}
T(s,t)\sim \beta(t)s^{\alpha(t)},
\qquad
s\to\infty,\quad t\ {\rm fixed},
\label{eq:regge-growth}
\end{equation}
with the leading trajectory satisfying $\alpha(0)\leq1$ in a theory compatible with unitarity. A resonance of spin $J$ and mass $M$ is exchanged in the kinematic region with non-zero transfer momentum, $t=M^2$, for which $\alpha(M^2)=J$.

This expectation should be contrasted with the simple asymptotic behavior of the primal ansatz $T^{\rm ans}$ in Eq.~\eqref{gamp}, which at any fixed truncation order tends to a constant plus $O(1/\sqrt{s})$ corrections. However, $T^{\rm ans}$ is a finite truncation of an expansion. As $N\to\infty$, it can still approximate a function with nontrivial Regge behavior over an increasingly large energy range.

\subsection{Regge theory in a nutshell}

Regge theory studies the analytic continuation of partial-wave coefficients in angular momentum $\ell$. Consider the partial-wave expansion of the amplitude in the crossed, $t$-channel,
\begin{equation}
T(s,t)
=
\sum_{\ell=0}^{\infty}
(2\ell+1) f_\ell(t) P_\ell(x),
\qquad
x=1+\frac{2s}{t-4},
\qquad
t\geq4,\quad s<0,\quad u<0 .
\label{eq:pwave_exp_t}
\end{equation}
The partial-wave coefficients can be analytically continued in $\ell$, and the sum can be rewritten via a Sommerfeld-Watson transform in the $\ell$-plane,
\begin{equation}
T(s,t)
=
\frac{1}{2i}
\int_{\mathcal C} d\ell\,
(2\ell+1)a(\ell,t)
\frac{P_\ell(-x)}{\sin(\pi\ell)} ,
\label{eq:SW}
\end{equation}
where the contour $\mathcal C$ encircles the non-negative even integers. The residues of the poles of $1/\sin(\pi\ell)$ reproduce the partial-wave expansion. Here $P_\ell(x)={}_2F_1(\ell+1,-\ell,1,(1-x)/2)$, and a useful analytic continuation of the partial waves is provided by the Froissart-Gribov representation,
\begin{equation}
a(\ell,t)
=
\frac{1}{\pi}
\int_{x_0}^{\infty} dx\,
{\rm Disc}_{s}\,T(s(x),t)\,Q_\ell(x),
\label{eq:FG}
\end{equation}
with $x_0=1+\frac{8}{t-4}$, $Q_\ell(x)$ is the Legendre  function of the second kind and $\text{Disc}_s f(s) \equiv (f(s+i\epsilon)-f(s-i\epsilon))/2i$. Froissart-Gribov representation agrees with the physical partial waves at even integer $\ell$ under the usual analyticity and boundedness assumptions.\footnote{
Naively, one could try to analytically continue the usual partial-wave integral directly. However, endpoint factors such as $(-1)^\ell=e^{i\pi\ell}$ obstruct the uniqueness properties required for the continuation. The Froissart-Gribov representation avoids this issue by expressing the continuation in terms of the crossed-channel discontinuity.
}

Assume that $a(\ell,t)$ is analytic and bounded in the half-plane ${\rm Re}\,\ell>\ell_0$, for some $\ell_0\in\mathbb{R}$. Deforming the contour to the vertical line ${\rm Re}\,\ell=\ell_0$ gives
\begin{equation}
T(s,t)
=
\sum_{\ell=0}^{\lfloor\ell_0\rfloor}
(2\ell+1)f_\ell(t)P_\ell(x)
+
\frac{1}{2i}
\int_{\ell_0-i\infty}^{\ell_0+i\infty}
d\ell\,
(2\ell+1)a(\ell,t)
\frac{P_\ell(-x)}{\sin(\pi\ell)} .
\label{eq:SW_deformed_1}
\end{equation}
We may then deform the contour further to the left. Suppose that $a(\ell,t)$ has only simple poles in the strip $\ell_1\leq{\rm Re}\,\ell\leq\ell_0$, located at $\ell=\alpha_i(t)$ with residues $\beta_i(t)$. Passing to the new contour on ${\rm Re}\,\ell=\ell_1$ yields
\begin{equation}
\begin{aligned}
T(s,t)
&=
\sum_{\ell=0}^{\lfloor\ell_1\rfloor}
(2\ell+1)f_\ell(t)P_\ell(x)
+
\frac{1}{2i}
\int_{\ell_1-i\infty}^{\ell_1+i\infty}
d\ell\,
(2\ell+1)a(\ell,t)
\frac{P_\ell(-x)}{\sin(\pi\ell)}
\\
&\quad
-
\sum_{i\in{\rm strip}}
\frac{(2\alpha_i(t)+1)\beta_i(t)P_{\alpha_i(t)}(-x)}
{\sin(\pi\alpha_i(t))}.
\end{aligned}
\label{eq:reggepole}
\end{equation}
The last term is the Regge-pole contribution.

The large-energy behavior in the non-crossed channel, corresponding to the $s$-channel Regge limit at fixed $t$, follows from Eq.~\eqref{eq:reggepole}. At large argument, $P_\alpha(-x)\sim(-x)^\alpha$, and in the $s$-channel Regge limit one has $-x\sim s$. Therefore the pole with largest real part dominates. Denoting the leading pole by $\alpha_p(t)$, one obtains
\begin{equation}
\lim_{\substack{s\to\infty\\ t\ {\rm fixed}}}
T(s,t)
\sim
\beta_p(t)\,s^{\alpha_p(t)} .
\label{eq:regge_growth_from_SW}
\end{equation}
This is the Regge behavior used in the main text.

A few remarks are useful:
\begin{itemize}
\item The Froissart bound on the forward amplitude implies $\alpha_p(0)\leq1$.
\item Equation~\eqref{eq:reggepole} develops a pole whenever $\alpha_i(t_\ell)=\ell\in2\mathbb{Z}_+$. This is interpreted as a spin-$\ell$ excitation exchanged in $t$-channel, with the complex position $t_\ell$ determining its mass and width. All resonances obtained from the same function $\alpha_i(t)$ lie on the same Regge trajectory.
\item The leading trajectory carries vacuum quantum numbers and controls the dominant high-energy behavior.
\end{itemize}

\subsection{Measurements of the asymptotic Regge limit}

In Eq.~\eqref{eq:regge_moments} we introduced Regge moments as diagnostics of the emerging high-energy behavior of the primal amplitudes. Their role is to detect whether an amplitude develops a regime compatible with Eq.~\eqref{eq:regge-growth} as the truncation order $N$ is increased.

Consider the two sample amplitudes discussed in Fig.~\ref{fig:regge_emergence}, one on the $AB$ arc and one on the $AC$ arc. We measure the first Regge moment $A_t(\Lambda)$ at several values of $N$ and compute the effective Regge intercept from its logarithmic derivative. In terms of the inverse-energy variable $y\equiv\Lambda^{-1/2}$, we define
\begin{equation}
A'_t(y)
\equiv
-\frac{1}{2}
\frac{\partial}{\partial\log y}
\log\left|A_t(1/y^2)\right| .
\label{eq:alpha_eff_y}
\end{equation}
This quantity coincides with $\alpha_{\rm eff}(t)$ in a regime where the Regge moment scales as a power of $\Lambda$.

\begin{figure}[t]
  \centering
  \includegraphics[width=0.48\linewidth]{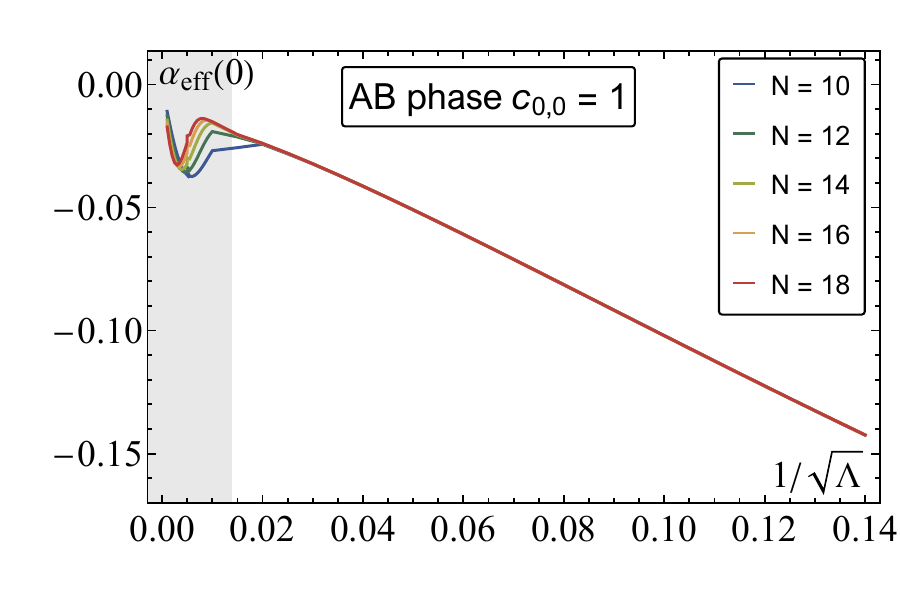}
  \hspace{.5cm}
  \includegraphics[width=0.455\linewidth]{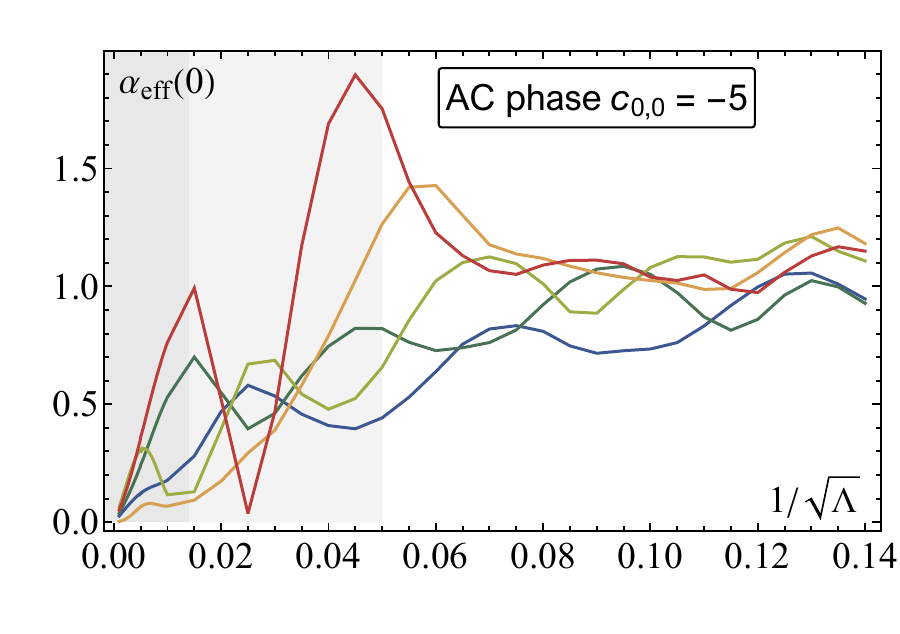}
  \caption{
  Extraction of the effective Regge intercept $\alpha_{\rm eff}(0)$ for the two amplitudes shown in Fig.~\ref{fig:regge_emergence}. The $AB$ example approaches marginal behavior, while the $AC$ example develops a stable plateau near $\alpha_{\rm eff}(0)\simeq1$ in the emergent Regge regime.
  }
  \label{fig:aeff_emergence}
\end{figure}

The results are shown in Fig.~\ref{fig:aeff_emergence}. As expected, the log-derivative converges in $N$ to a stable function inside the emergent Regge window. For the $AC$ example, corresponding to $c_{0,0}=-5$, a clear plateau develops,
\begin{equation}
A'_0(y)\simeq1,
\qquad
0.05\lesssim y\lesssim0.14 \quad(50 \lesssim \Lambda \lesssim 400 ),
\end{equation}
consistent with Regge growth controlled by a leading trajectory with intercept $\alpha_p(0)\simeq1$. By contrast, the $AB$ example does not display a positive-intercept plateau. Instead, the effective intercept approaches zero, in agreement with the marginal Regge behavior discussed in the main text. See also App.~\ref{appABphase} for more details on the decaying behavior.

There is however an important caveat in this numerical exercise: Extracting asymptotic behavior from a finite-$N$ ansatz requires an energy window that is high compared with the intrinsic scales of the amplitude, but not so high that the built-in finite-$N$ asymptotics of the $\rho$-ansatz dominates. This window changes substantially along the boundary. For example, along the $AC$ arc the relevant scale is controlled by the lightest spin-two resonance. As shown in Fig.~\ref{fig:AC_cross_sections}, its mass moves from threshold at the $C$ cusp to arbitrarily high energy as the free theory point $A$ is approached. An energy that is already asymptotic near $C$ may still be below the higher-spin scale closer to $A$, and therefore not high enough to detect the emerging behavior.

\begin{figure}[t!]
    \centering
    \includegraphics[scale=0.5]{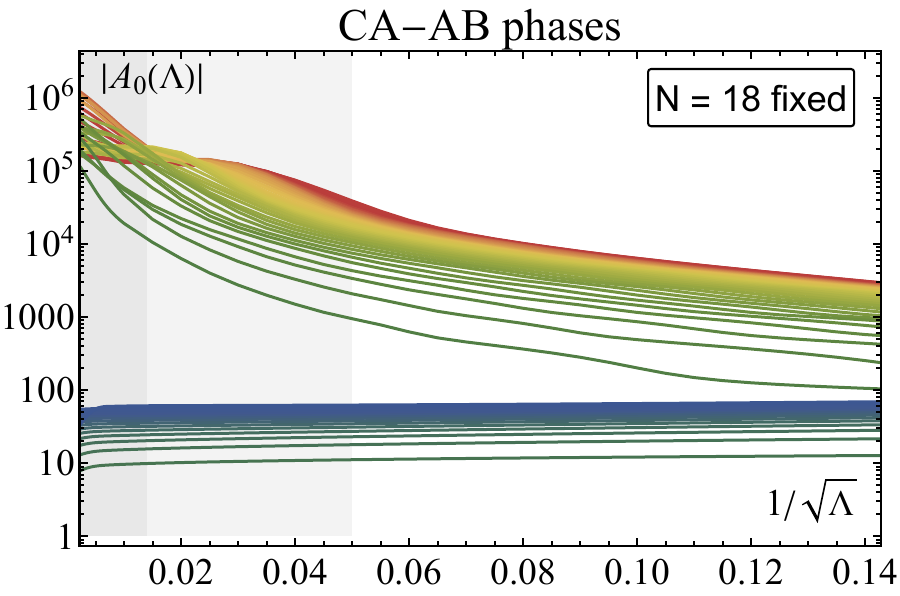}
    \includegraphics[scale=0.5]{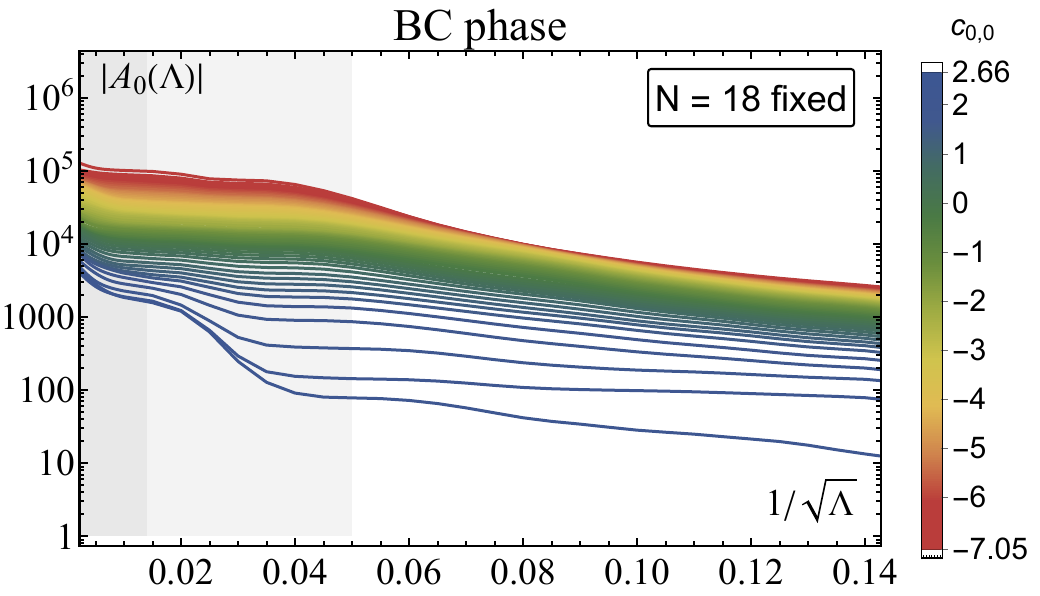} \\
    \includegraphics[scale=0.5]{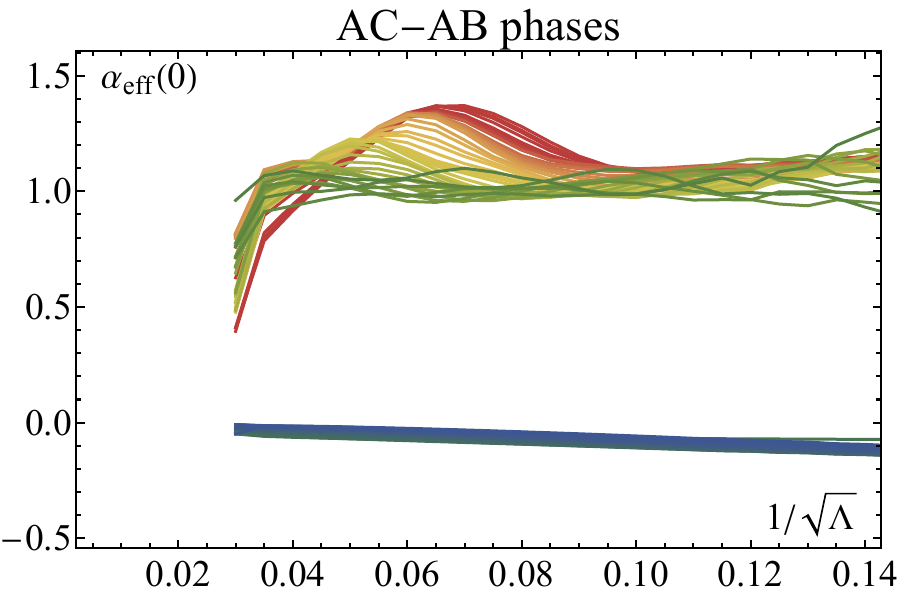}
    \includegraphics[scale=0.5]{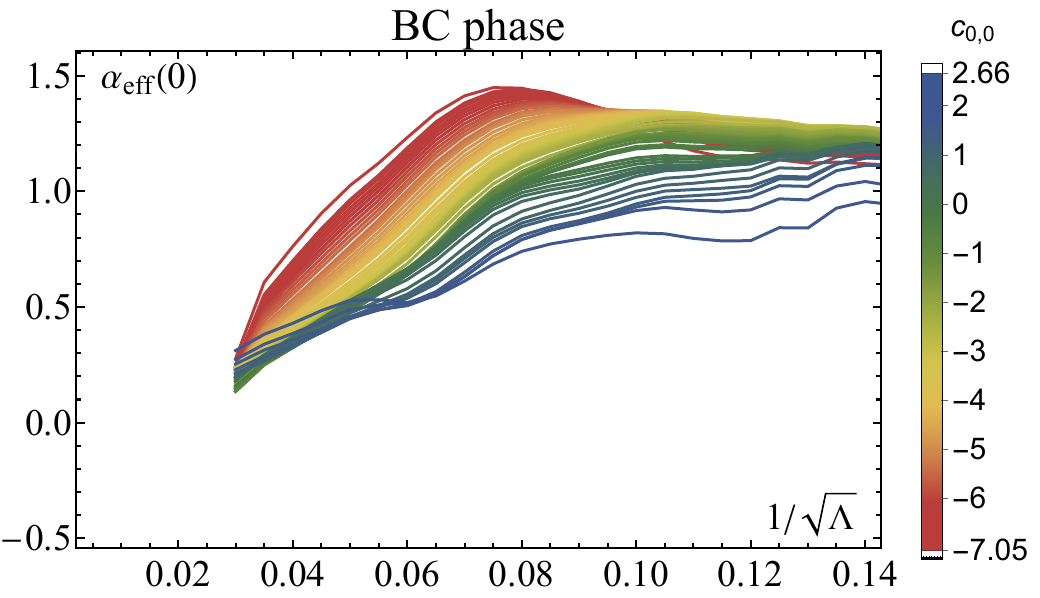}
    \caption{
    Regge moments and intercepts along the phases of Fig.~\ref{fig:dual_vs_primal} at a fixed truncation order $N=18$. \textbf{Upper panels:} Magnitude of the Regge moment as a function of the contour radius $\Lambda$ and $c_{0,0}$. \textbf{Lower panels:} Corresponding effective intercepts obtained from their logarithmic derivative. Stable plateaux indicate emergent Regge regimes, while non-monotonic or rapidly varying regions signal finite-$N$ contamination or insufficient separation from the intrinsic resonance scale.
    }
    \label{fig:alphaeff}
\end{figure}

This scale dependence is visible in Fig.~\ref{fig:alphaeff}. Stable plateaux in $\alpha_\text{eff}(0)$ plots indicate energy regions where a Regge exponent can be reliably extracted. Non-monotonic or rapidly varying regions signal either insufficient separation from the intrinsic resonance scale or contamination from the ultimate finite-$N$ asymptotic tail of the ansatz.

\section{Numerical analysis}
\label{app:numanal}

The extremal amplitudes obtained from the bootstrap contain more information than the low-energy coefficients shown in Fig.~\ref{fig:dual_vs_primal}. In this appendix we summarize numerical observables that support the phase structure described in the main text: zeros of partial waves, virtual states, resonance poles, Regge trajectories, and threshold couplings.

\subsection{Tracking zeros and resonances}

A useful diagnostic is the spectrum of zeros of the partial-wave $S$-matrices. With our convention
$S_\ell(s)=1+i\rho(s)f_\ell(s)$, a zero of $S_\ell(s)$ on the physical sheet corresponds, after analytic continuation through the elastic cut, to a pole on the second sheet. Real zeros in the interval $0<s<4$ are interpreted as virtual states, while complex zeros above threshold give resonance poles.

We locate zeros of $S_\ell(s)$ numerically using Newton's method,
\begin{equation}
    z_{n+1}
    =
    z_n
    -
    \frac{S_\ell(z_n)}{S_\ell'(z_n)} .
\end{equation}
The branch cuts make the problem mildly delicate. In practice, we first identify a zero for a given extremal amplitude and then use it as a hot start for neighboring amplitudes along the boundary. This allows us to track the trajectories continuously.

\begin{figure}[h!]
    \centering
    \includegraphics[width=0.8\linewidth]{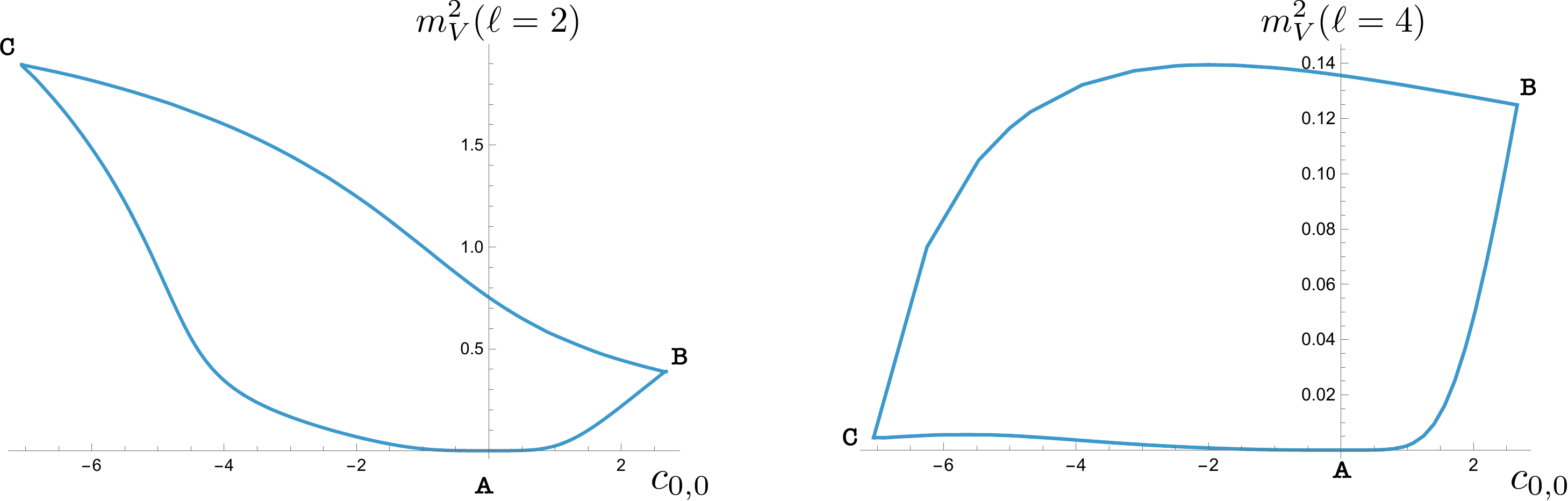}
    \caption{
    Virtual-state trajectories along the boundary.
    Left: spin $\ell=2$.
    Right: spin $\ell=4$.
    The plotted quantity is the position of the real zero of $S_\ell(s)$ in the interval $0<s<4$.
    }
    \label{fig:virtual_fun}
\end{figure}

Figure~\ref{fig:virtual_fun} shows two examples. The left panel tracks the spin-two virtual state, corresponding to a real zero of $S_2(s)$ in the interval $0<s<4$. Starting from the $C$ cusp, the zero moves toward the left-cut branch point at $s=0$ as one approaches the free theory. Continuing around the boundary, it re-emerges and evolves toward the $B$ cusp. The right panel shows the analogous trajectory for spin four. These virtual states provide a convenient global diagnostic because they can be followed continuously along the boundary, even in regions where resonances are absent or harder to extract.

Virtual states are particularly useful because they are present along the entire boundary. Resonances, by contrast, disappear along the $AB$ arc and re-emerge only at the threshold cusps. For spin four virtual state, shown in the right panel of Fig.~\ref{fig:virtual_fun}, the hierarchy between the two cusps is reversed: the mass at the $B$ cusp, $s\simeq0.125$, is larger than the mass at the $C$ cusp, $m_V^2\simeq4.5\times10^{-3}$.

\subsection{Threshold states and the three arcs}

\begin{figure}[h!]
    \centering
    \includegraphics[width=\linewidth]{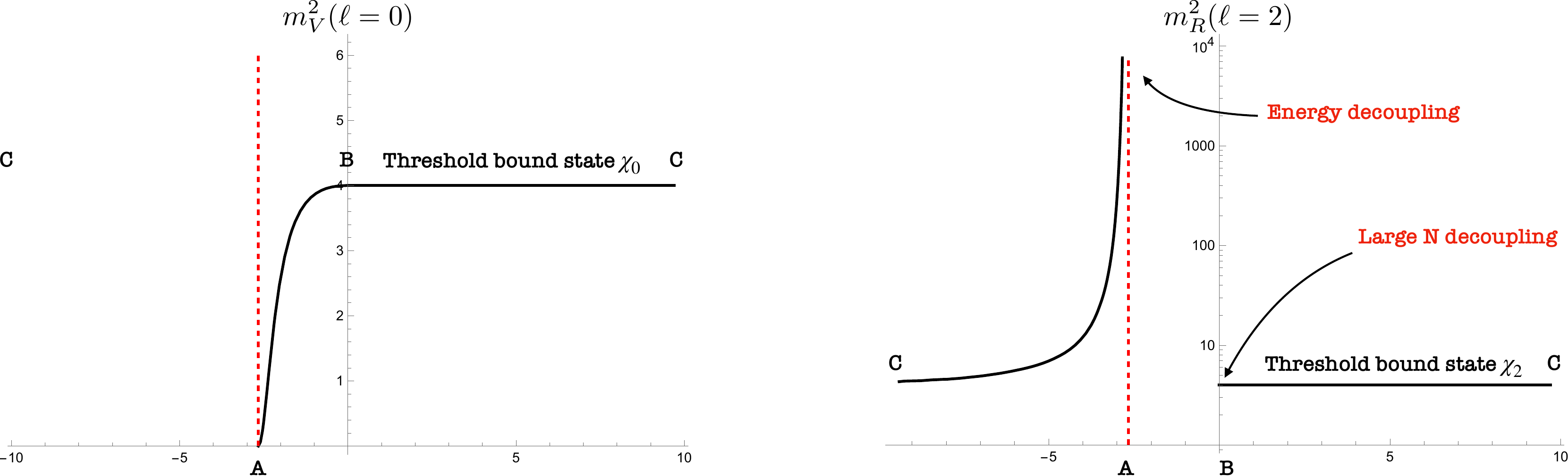}
    \caption{
    Threshold diagnostics along the boundary.
    The horizontal coordinate measures the distance from the cusp value $c_{0,0}^{\max}\simeq2.66$, with the $AC$ and $BC$ branches parameterized according to their natural orientation along the boundary ($c_{0,0}-\max c_{0,0,}$ on the lower boundary of the island, and $\max c_{0,0}-c_{0,0}$ on the upper boundary).
    Left: motion of the scalar virtual state and its continuation to the scalar threshold state.
    Right: motion of the lightest spin-two resonance, which becomes the spin-two threshold state at the $C$ cusp and decouples in coupling along the $BC$ arc.
    }
    \label{fig:chi0_and_chi2}
\end{figure}

The lightest scalar virtual state and the lightest spin-two resonance capture the two threshold mechanisms discussed in the main text. In Fig.~\ref{fig:chi0_and_chi2} we track these two states along the boundary.

In the left panel of Fig.~\ref{fig:chi0_and_chi2}, the scalar virtual state emerges from the left-hand cut and becomes a threshold bound state at the $B$ cusp. This state, denoted $\chi_0$, remains at threshold along the $BC$ arc until it is removed by the zero-pole cancellation described in Appendix~\ref{app:cdd}.

The right panel of Fig.~\ref{fig:chi0_and_chi2} shows the corresponding behavior of the lightest spin-two resonance. As the free point is approached along the $AC$ arc, its mass grows and the state decouples in energy. Moving toward the $C$ cusp, the resonance becomes lighter until it reaches threshold and becomes a stable spin-two bound state $\chi_2$. Continuing along the $BC$ arc toward $B$, its mass remains fixed while its coupling decreases and eventually vanishes.

\subsection{Higher-spin resonances and Regge trajectories}

\begin{figure}[h!]
    \centering
    \includegraphics[width=\linewidth]{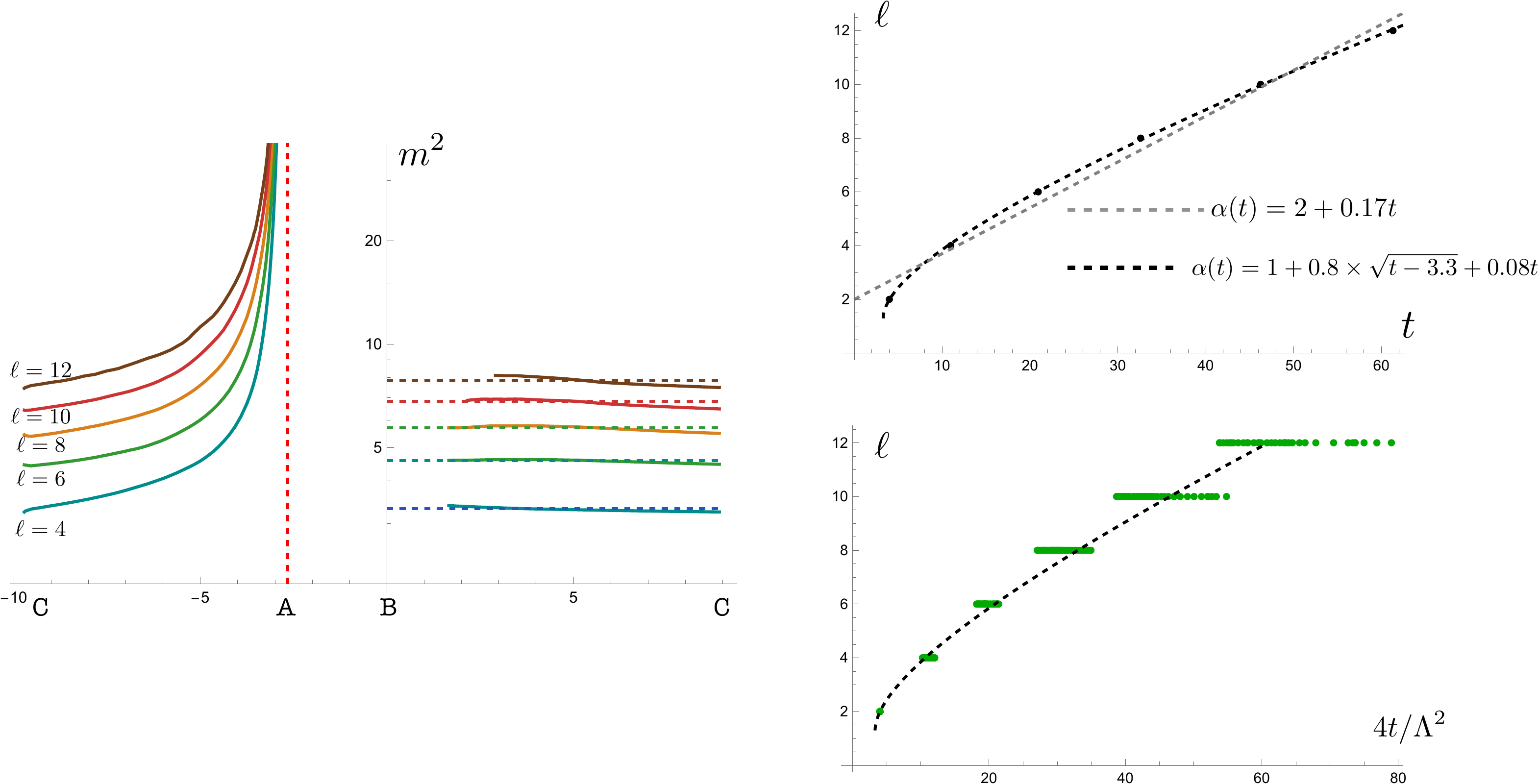}
    \caption{
    Higher-spin spectrum and leading Regge trajectory.
    Left: masses of higher-spin resonances along the boundary. The spectrum is pushed to high energy along the $AC$ arc, while remaining approximately fixed along $BC$.
    Top right: fit of the leading Regge trajectory using the phenomenological form $\alpha(t)=1+a\sqrt{t-d}+bt$.
    Bottom right: leading trajectory rescaled by the spin-two mass, showing the approximate self-similarity of the $AC$ energy-decoupling regime.
    }
    \label{fig:Regge_plots}
\end{figure}

We next examine the higher-spin resonances on the leading Regge trajectory. Figure~\ref{fig:Regge_plots} tracks their masses and the associated trajectory along the boundary.

Similarly to the spin-two state, the masses of higher-spin resonances increase as the free point is approached along $AC$, indicating decoupling in energy. Along the $BC$ arc, instead, these states remain approximately at fixed mass while their residues decrease. This is the large-$N_c$-like behavior described in the main text.

The top-right panel of Fig.~\ref{fig:Regge_plots} shows a fit to the leading trajectory. The best fit is not linear, but is well described by the phenomenological form
\begin{equation}
    \alpha(t)
    =
    1
    +
    a\sqrt{t-d}
    +
    b\,t .
\end{equation}
Such a square-root behavior is motivated by compatibility with Froissart growth, which allows Regge trajectories to develop square-root branch points at small $t$; see for instance Ref.~\cite{Correia:2025uvc}. Here we simply use this functional form as a fitting ansatz, leaving a deeper theoretical interpretation for future work.

The bottom-right panel of Fig.~\ref{fig:Regge_plots} shows the higher-spin masses rescaled by the mass of the leading spin-two resonance. The low-spin part of the trajectory remains approximately stable under this rescaling, supporting the self-similar energy-decoupling picture of the $AC$ arc. Deviations become more visible at higher spin and close to the free point, where numerical convergence is slower.

\begin{figure}[h!]
    \centering
    \includegraphics[scale=0.5]{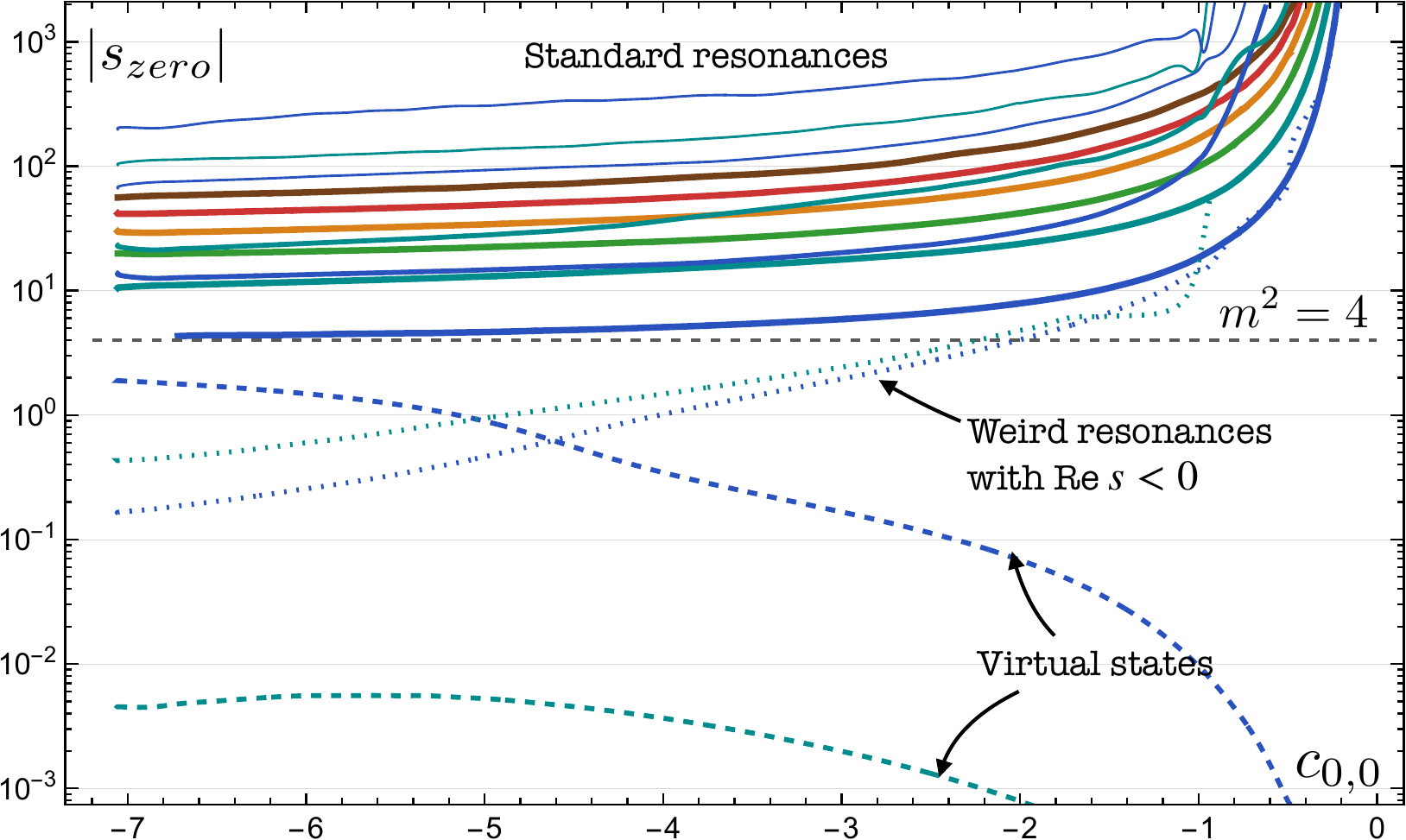}
    \caption{
    Full spectrum of zeros tracked along the boundary. Solid curves denote resonance trajectories, while dashed curves denote virtual states. The plot gives a global view of the spectrum underlying the three phases discussed in the main text.
    }
    \label{fig:full_spectrum}
\end{figure}

Figure~\ref{fig:full_spectrum} provides a global view of the resonance dynamics along the $AC$ arc. The thick blue curve denotes the leading spin-two resonance, while thinner blue curves correspond to heavier spin-two resonances. The same plot also shows daughter trajectories for spin four and spin six. Dashed curves near the bottom correspond to higher-spin virtual states, while dotted curves denote additional zeros with negative real part. These latter trajectories appear systematically in the numerical bootstrap solutions, although their physical interpretation is less clear. As the free point is approached, the plot shows the collective energy decoupling of the spectrum: all resonance trajectories are pushed to higher energies, with the negative-real-part zeros decoupling only at the latest stages.

\subsection{Low-energy constants and threshold diagnostics}
\label{app:low_energy_constants}

Beyond the spectrum, the extremal amplitudes provide access to threshold low-energy constants and effective couplings. These quantities give useful diagnostics of the mechanisms described in the main text, in particular the scalar zero-pole cancellation and the large-$N_c$-like weakening of higher-spin resonances along $BC$.

\begin{figure}[h!]
    \centering
    \includegraphics[width=0.8\linewidth]{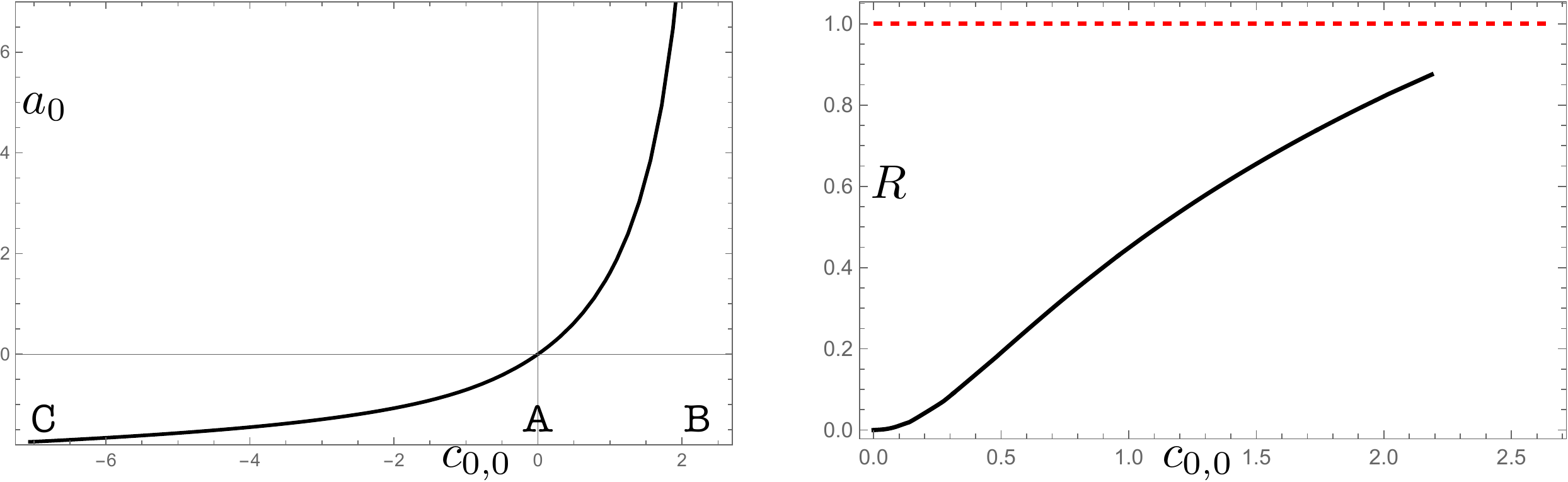}
    \caption{
    Left: S-wave scattering length $a_0$ as function of $c_{0,0}$ on the lower boundary of the island. Right: ratio R \eqref{eq:ratio_R} used to diagnose the universal behavior of the scattering length as we approach the B cusp.
    }
    \label{fig:spin0_scatt_length}
\end{figure}

In the left panel of Fig.~\ref{fig:spin0_scatt_length}, we follow the scalar scattering length $a_0$ along the lower boundary of the island. It starts from its minimum value at the $C$ cusp, compatible with Refs.~\cite{Lopez:1974cq,Paper3}, and diverges as $c_{0,0}$ approaches its maximal value. This divergence is the characteristic signal of a virtual state approaching threshold. The CDD model of Appendix~\ref{app:cdd} predicts the universal scaling
\begin{equation}
    4-m_V^2\simeq \frac{4}{a_0^2}.
\end{equation}
We therefore consider the ratio
\begin{equation}
    R=\frac{4-m_V^2}{4/a_0^2}.
\label{eq:ratio_R}
\end{equation}
The right panel of Fig.~\ref{fig:spin0_scatt_length} shows this ratio for positive values of $c_{0,0}$. It approaches one near threshold, as expected, although this regime is numerically delicate because convergence becomes slower close to the cusp.

\begin{figure}[h!]
    \centering
    \includegraphics[width=0.8\linewidth]{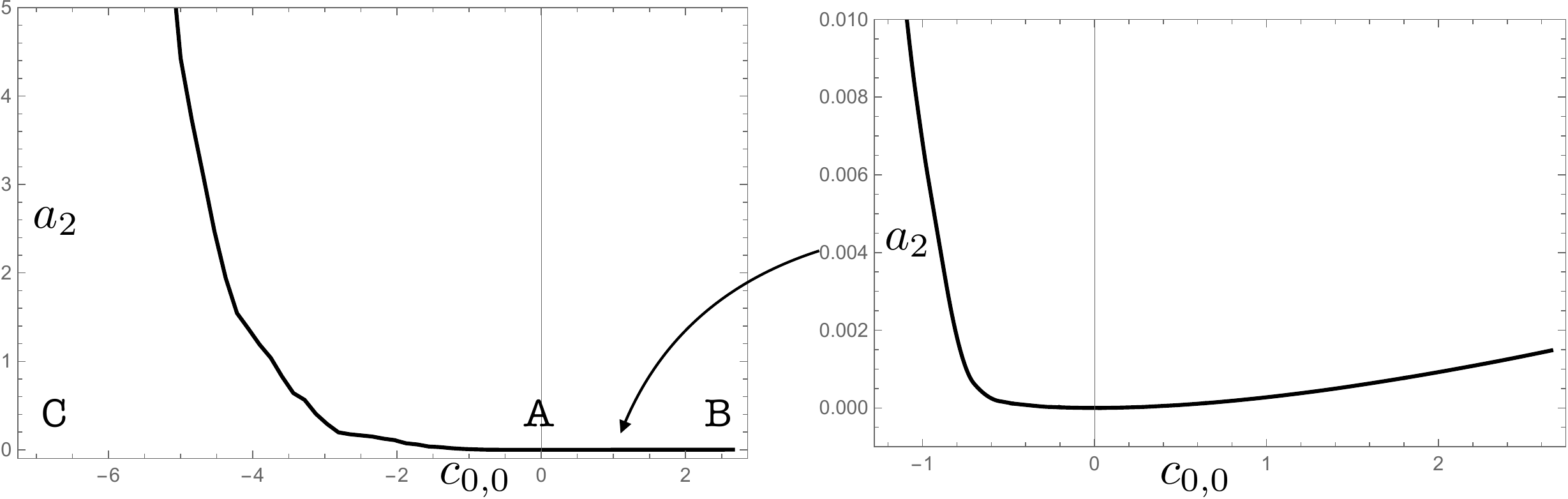}
    \caption{
    Spin-two scattering length $a_2$ as a function of $c_{0,0}$ along the lower boundary of the island. Left: divergence of $a_2$ as the $C$ cusp is approached, signalling the spin-two resonance reaching threshold. Right: zoom near the free theory, showing that $a_2$ has a minimum at $c_{0,0}=0$ and remains positive.
    }
    \label{fig:spin2_scatt_length}
\end{figure}

In the left panel of Fig.~\ref{fig:spin2_scatt_length}, we show the analogous behavior for the spin-two scattering length $a_2$. Its divergence as the $C$ cusp is approached confirms that a spin-two resonance reaches threshold and becomes the spin-two bound state $\chi_2$. This behavior follows from the simple pole model discussed in Appendix~\ref{sec:K-matrix},
\begin{equation}
    f_2(s)
    \simeq
    \frac{g^2}{80\pi}\frac{(s-4)^2}{s_R-s}
    =
    \frac{g^2}{80\pi}\frac{\epsilon^2}{\Delta-\epsilon},
\end{equation}
where $\epsilon=s-4$ and $\Delta=s_R-4$. Taking $\epsilon\ll\Delta\ll1$ and matching to $f_2(s)\sim a_2(s-4)^2$ gives
\begin{equation}
    a_2\sim \frac{g^2}{80\pi}\frac{1}{s_R-4}.
\end{equation}
Thus the divergence of $a_2$ measures the approach of the spin-two pole to threshold.

The right panel of Fig.~\ref{fig:spin2_scatt_length} zooms in near the free theory. The coefficient $a_2$ has a minimum at $c_{0,0}=0$ and is positive on both sides. Its value near the free theory can be computed perturbatively from the one-loop amplitude of Appendix~\ref{app:one-loop}. With the convention $f_2(s)=2a_2(s-4)^2+\cdots$, the relevant sum rule is
\begin{equation}
    a_2
    =
    \frac{1}{480\pi^2}
    \int_4^\infty ds\,
    \frac{\Im T(s,4)}{s^3}.
\end{equation}
Using the one-loop result for $\Im T(s,4)$, this gives
\begin{equation}
    a_2^{\rm 1-loop}
    =
    \frac{c_{0,0}^2}{15\pi}
    \int_4^\infty ds\,
    \frac{\sqrt{s-4}}{s^{7/2}}
    =
    \frac{c_{0,0}^2}{900\pi},
\end{equation}
in excellent agreement with the numerical data.

\begin{figure}[h!]
    \centering
    \includegraphics[width=0.9\linewidth]{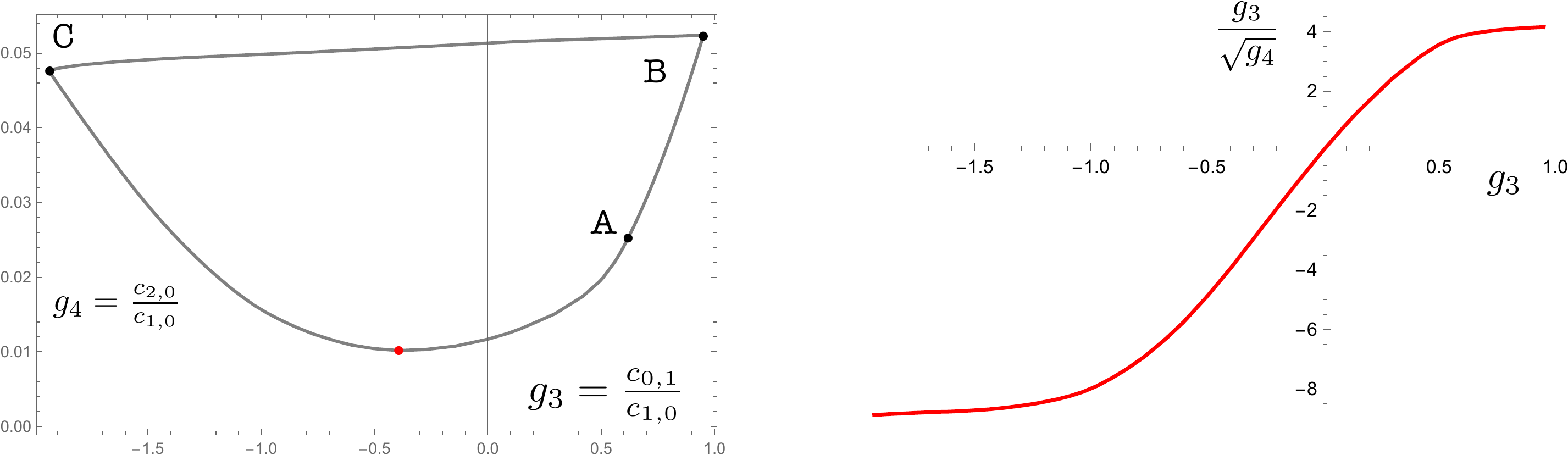}
    \caption{
    Ratios of low-energy S-matrix data. Left: the boundary in the $(g_3,g_4)$ plane, with the two cusps associated with the scalar and spin-two threshold states. The red point denotes the minimum of $g_4$. Right: plot of $g_3/\sqrt{g_4}$ to measure how well the boundary is approximated by a parabola.
    }
    \label{fig:g3g4plots}
\end{figure}

Threshold constants often diverge and can become numerically unstable. It is therefore useful to also consider ratios of low-energy S-matrix data. In the left panel of Fig.~\ref{fig:g3g4plots}, we show
\begin{equation}
    g_3=\frac{c_{0,1}}{c_{1,0}},
    \qquad
    g_4=\frac{c_{2,0}}{c_{1,0}} .
\end{equation}
The plot reveals two cusps associated with $B$ and $C$, while the free theory is reached smoothly along the boundary.

The value at the $B$ cusp can be predicted from the scalar threshold model. The appropriate endpoint singularity is the threshold square-root behavior, for which $\Im T(s,t)\propto 1/\sqrt{s-4}$ at fixed $t=4/3$. Using the sum rules of Appendix~\ref{sec:3dshape}, one finds
\begin{equation}
    g_3=\frac{15}{16}\simeq0.94,
    \qquad
    g_4=\frac{105}{2048}\simeq0.05,
\end{equation}
in precise agreement with the numerical result.

The $BC$ arc corresponds to the nearly flat upper boundary of the figure, where $g_4$ is approximately constant. This has a simple interpretation: if the higher-spin spectrum is approximated by a sum of poles at fixed masses, then the even ratio $g_4$ is controlled mainly by these masses and is largely insensitive to an overall suppression of residues. This is consistent with the large-$N_c$-like decoupling mechanism discussed in the main text. The agreement is not expected to be exact, because the same amplitudes also contain the scalar threshold state $\chi_0$.

The right panel of Fig.~\ref{fig:g3g4plots} shows that the lower boundary is well approximated by a parabola near the threshold cusps. Near $C$, the imaginary part is dominated by the spin-two resonance approaching threshold. Neglecting the angular dependence of the residue, the same threshold scale controls both ratios: $g_3$ carries one inverse power of this scale, while $g_4$ carries two. This gives the dimensional estimate $g_4\propto g_3^2$. The same logic applies near the $B$ cusp, where the relevant scale is instead the distance of the scalar CDD zero/virtual state from threshold, provided the higher-spin contributions are neglected.

Finally, the red point denotes the minimum value of $g_4$ along the boundary. We suspect that this point coincides with a \emph{prevertex} of the higher-dimensional amplitude space; see Appendix~\ref{sec:3dshape}.

\begin{figure}[t!]
    \centering
    \includegraphics[width=0.9\linewidth]{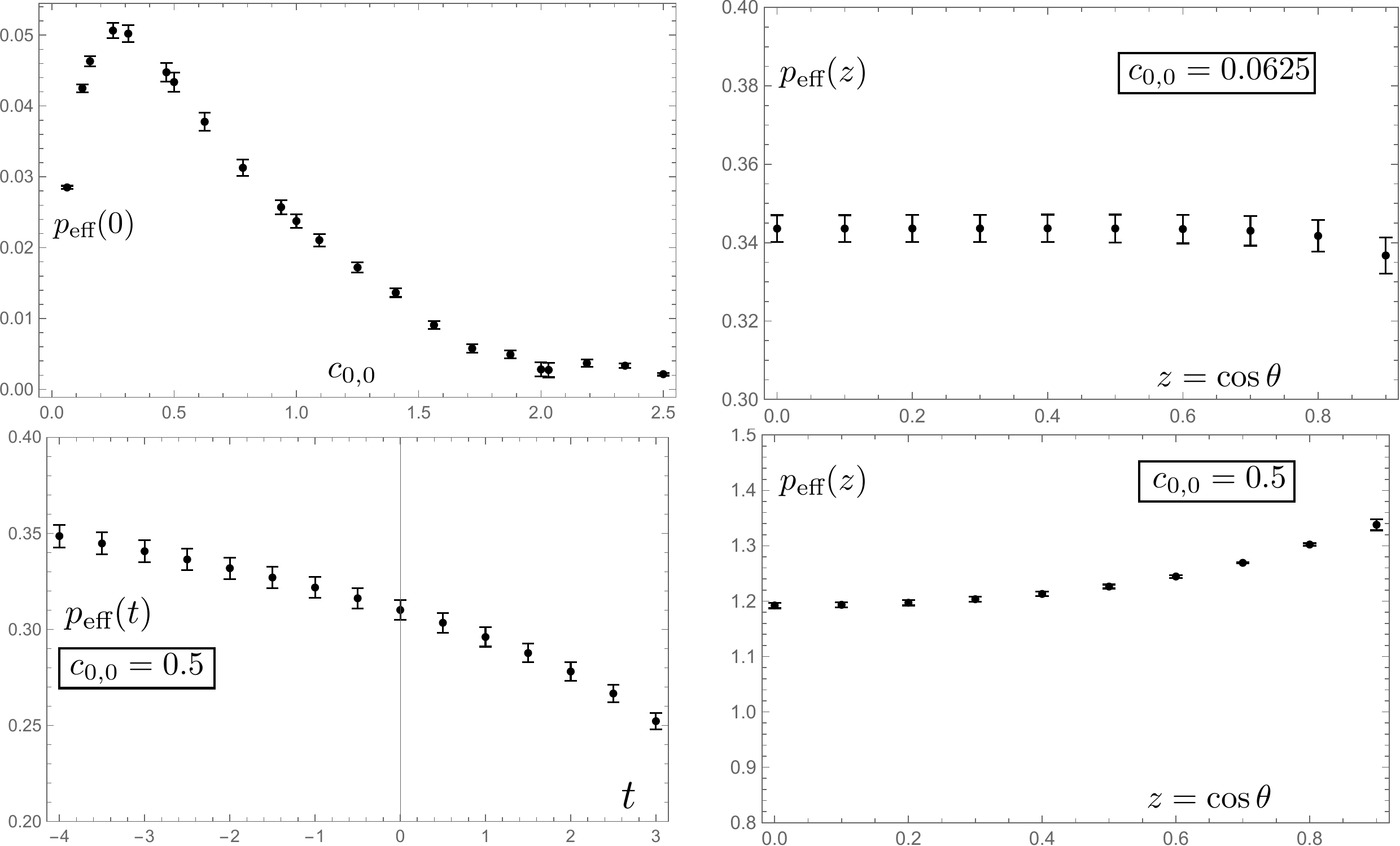}
    \caption{
    Effective logarithmic exponent $p_{\rm eff}$ extracted from high-energy fits on the $AB$ arc.
    Top left: $p_{\rm eff}$ at fixed $t=0$ as a function of $c_{0,0}$, showing a nontrivial dependence and a tendency to decrease toward zero near the edges of the allowed region.
    Top right: $p_{\rm eff}(z)$ at fixed $c_{0,0}=0.0625$, remaining approximately constant as a function of the scattering angle $z=\cos\theta$.
    Bottom left: $p_{\rm eff}(t)$ at fixed $c_{0,0}=0.5$, displaying a smooth dependence on the momentum transfer.
    Bottom right: $p_{\rm eff}(z)$ at fixed $c_{0,0}=0.5$, showing a mild but systematic angular dependence.
    Error bars are obtained from the variation over fitting windows and provide a conservative estimate of uncertainties.
    }
    \label{fig:AB_log_final}
\end{figure}

\subsection{AB phase}
\label{appABphase}

We analyze the high-energy behavior of amplitudes on the $AB$ arc both at fixed momentum transfer and at fixed scattering angle. We define
\begin{equation}
f(s)
\equiv
-\log |T(s,t)|,
\qquad
g(s)
=
-\log
\left|
T\left(s,\frac{4-s}{2}(1-z)\right)
\right|,
\end{equation}
where $z=\cos\theta$. We fit these functions over sliding windows in $\log s$ and $\log\log s$, starting from $s=200$ and restricting to the region where the amplitude is numerically converged. The fitting ansatzes are
\begin{equation}
y(s)\simeq a+c_{\rm eff}\log s,
\qquad
y(s)\simeq a+p_{\rm eff}\log\log s,
\end{equation}
which define effective power-law and logarithmic exponents. Scanning over window sizes $w$ gives $p_{\rm eff}(w)$, from which we estimate
\begin{equation}
p_{\rm eff}
=
\frac{1}{N_{\mathcal W}}
\sum_{w\in\mathcal W}p_{\rm eff}(w),
\qquad
\Delta p_{\rm eff}
=
\frac{1}{2}
\left(
\max_{\mathcal W}p_{\rm eff}
-
\min_{\mathcal W}p_{\rm eff}
\right).
\end{equation}
This provides a conservative estimate of the uncertainty associated with the choice of fitting window.

Power-law fits do not stabilize over the accessible range, whereas logarithmic fits are more robust. At fixed $c_{0,0}$, the fixed-angle amplitude is consistent with
\begin{equation}
|T(s,t)|
\sim
(\log s)^{-p_{\rm eff}(t)},
\qquad
|T(s,t(s,z))|
\sim
(\log s)^{-p_{\rm eff}(z)} ,
\end{equation}
with a positive exponent that varies smoothly with $t$ and more mildly with the scattering angle. Conversely, at fixed $t=0$, $p_{\rm eff}=p_{\rm eff}(c_{0,0})$ exhibits a nontrivial dependence on the position along the arc, approaching zero near both the free theory and the strongly coupled endpoint. This is consistent with the fact that the amplitude must eventually leave the logarithmic regime outside the accessible numerical window. The resulting effective logarithmic exponents are shown in Fig.~\ref{fig:AB_log_final}. They support the picture that the $AB$ arc is marginal in the Regge limit: no stable positive power-law growth is observed, while logarithmic fits remain robust over the accessible numerical window.

\begin{figure}[h!]
    \centering
    \includegraphics[width=0.9\linewidth]{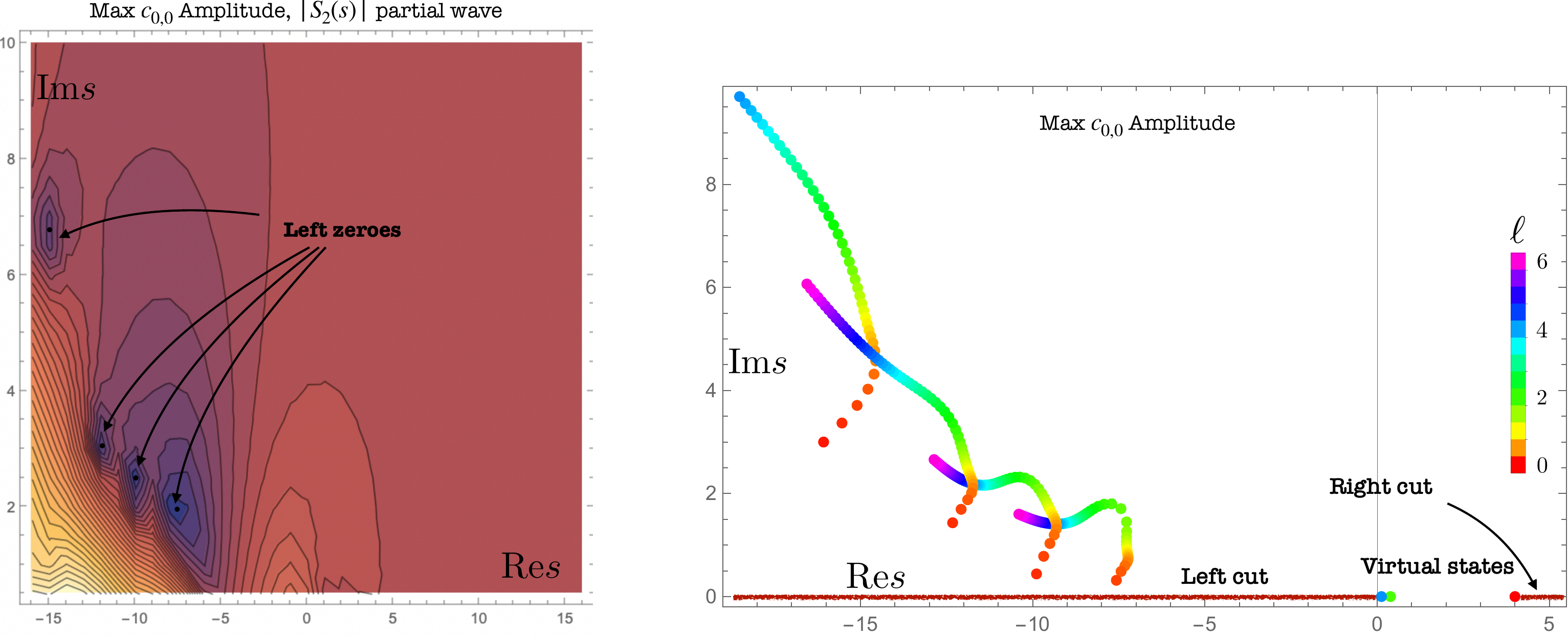}
    \caption{
    Left: density plot of $|S_2(s)|$ in the complex $s$ plane. Black dots denote zeros.
    Right: analytic continuation in spin of the zeros of $S_2$, showing that they organize into Regge trajectories.
    }
    \label{fig:max_coupling_spectrum}
\end{figure}

Beyond the virtual states, amplitudes on this arc do not exhibit conventional massive resonances at $\mathrm{Re}\,s>0$. They do, however, contain many resonance trajectories with $\mathrm{Re}\,s<0$. In the left panel of Fig.~\ref{fig:max_coupling_spectrum}, we show a density plot of the partial wave $|S_2(s)|$ for the amplitude at the $B$ cusp. Red denotes the region where $|S_2|\sim1$, yellow larger values, and blue smaller values. The black dots correspond to zeros on the physical sheet, and hence to poles on the second sheet.

For $\mathrm{Re}\,s>0$, the amplitude remains of order one and does not show any pronounced resonant structure. By contrast, for $\mathrm{Re}\,s<0$ it becomes large near the left-hand cut. Between these two regions there is a transition zone populated by many zeros, separating the left cut from the order-one region. Similar structures have been reported in Refs.~\cite{Guerrieri:2022sod,Guerrieri:2023qbg,Correia:2025uvc}. Their systematic appearance calls for a deeper explanation, which is presently lacking.

The right panel of Fig.~\ref{fig:max_coupling_spectrum} shows that these negative-real-part resonances organize into Regge trajectories. Upon analytic continuation in spin, the corresponding zeros move continuously and can be followed from spin zero up to spin six.

\subsection{AC phase}

\begin{figure}[h!]
    \centering
    \includegraphics[width=0.9\linewidth]{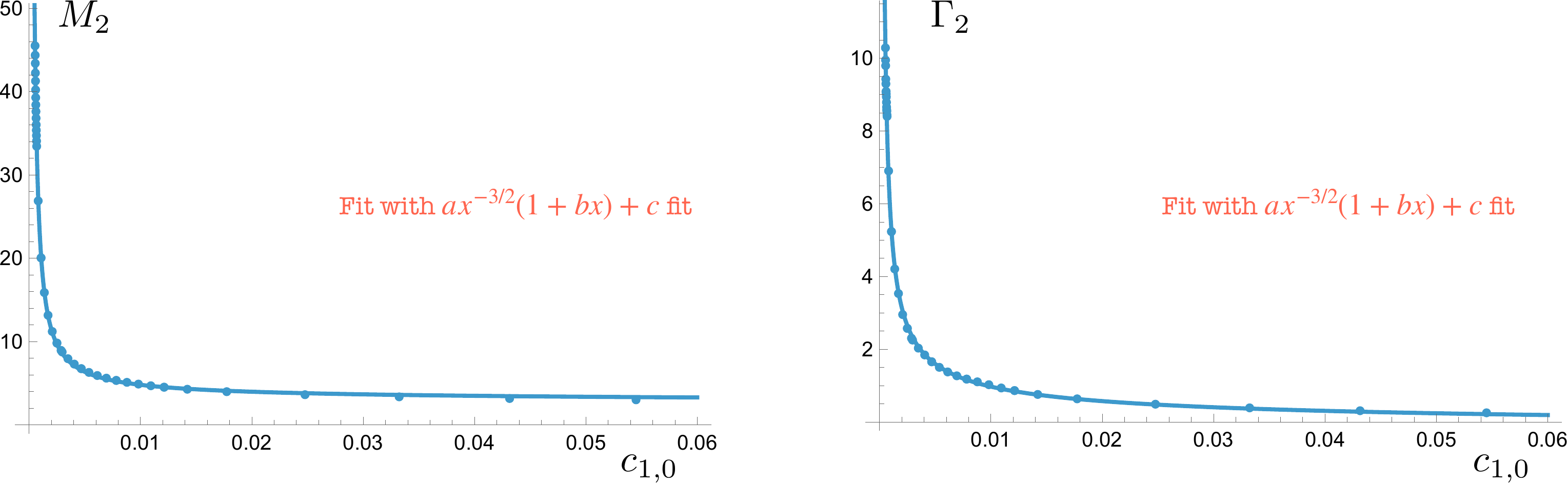}
    \caption{
    Fits to the mass and width of the lightest spin-two resonance on the $AC$ arc using a $c_{1,0}^{-3/2}$ scaling. Both quantities grow rapidly as the free point is approached, supporting the energy-decoupling picture of the higher-spin sector. The fits are indicative and do not include a full extrapolation in truncation order.
    }
    \label{fig:fit_masses_AC}
\end{figure}

In Fig.~\ref{fig:fit_masses_AC} we show a simple power-law fit of the mass $M_2$ and width $\Gamma_2$ of the lightest spin-two resonance on the $AC$ arc as functions of the low-energy coefficient $c_{1,0}$. The fit is intended as an indicative diagnostic rather than a precision determination. It shows that, as the free point is approached, both the mass and the width grow rapidly. Among the simple power laws we tested, the scaling $c_{1,0}^{-3/2}$ gives the most consistent description of the data. This supports the energy-decoupling picture described in the main text: the higher-spin sector is pushed to high energy while remaining strongly coupled at its own characteristic scale.

This behavior should not be interpreted as a single weakly coupled pole saturating the $c_{1,0}$ sum rule. Rather, the observed scaling is evidence for a collective motion of the higher-spin Regge sector. The width grows together with the mass, with no evidence that $\Gamma_2/M_2$ tends to zero. For the spin-two state we find $\Gamma_2/M_2\sim0.15$, suggestively close to the QCD value $\Gamma_\rho/M_\rho\sim0.2$ for the $\rho$ meson. This supports the intuition that, on the $AC$ arc, the spectrum remains strongly coupled. Thus the higher-spin sector decouples from the low-energy scattering particles by being pushed to high energy, rather than by becoming narrow.

The numerical systematics of this fit should be treated with care. Convergence in $N$ and $L$ is not uniform along the $AC$ arc because the higher-spin scale itself is moving. In particular, the width is observed empirically to converge more slowly than the mass.

\subsection{BC phase}

\begin{figure}[h!]
    \centering
    \includegraphics[width=0.45\linewidth]{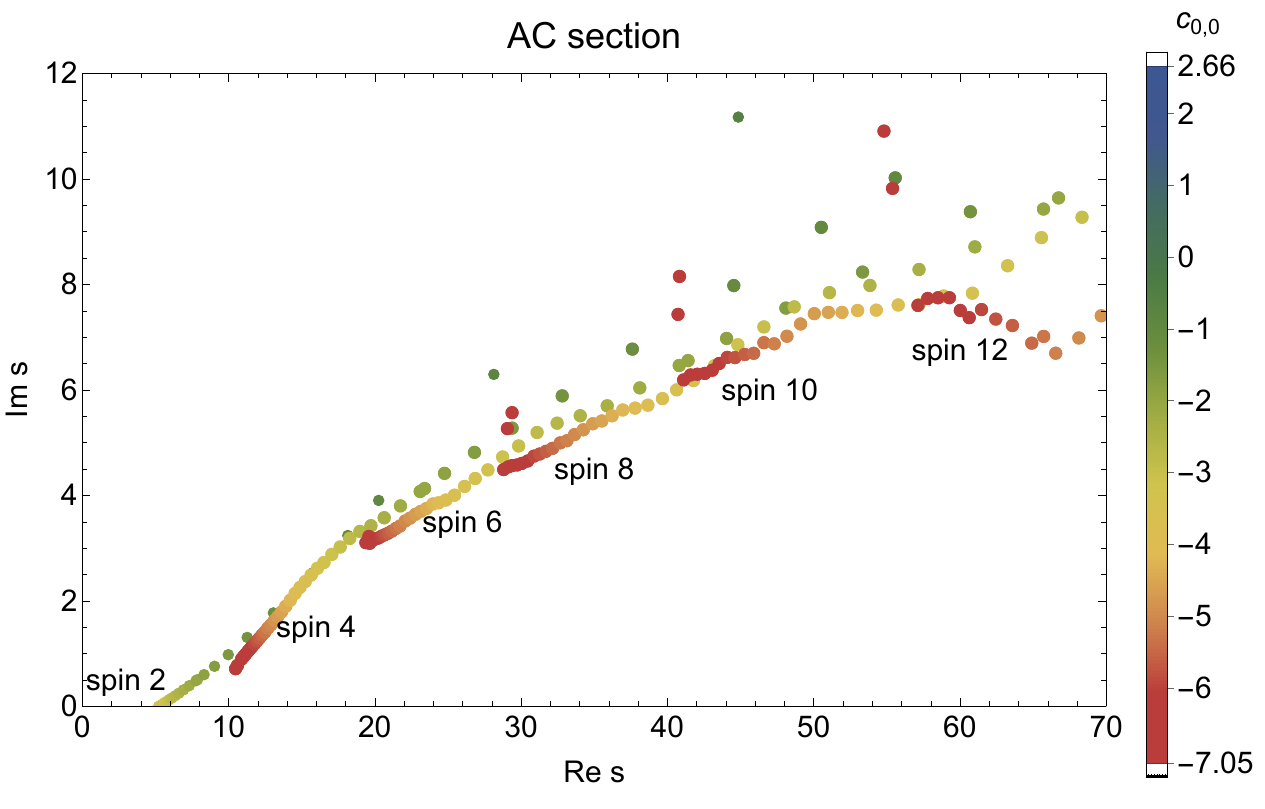}
    \includegraphics[width=0.45\linewidth]{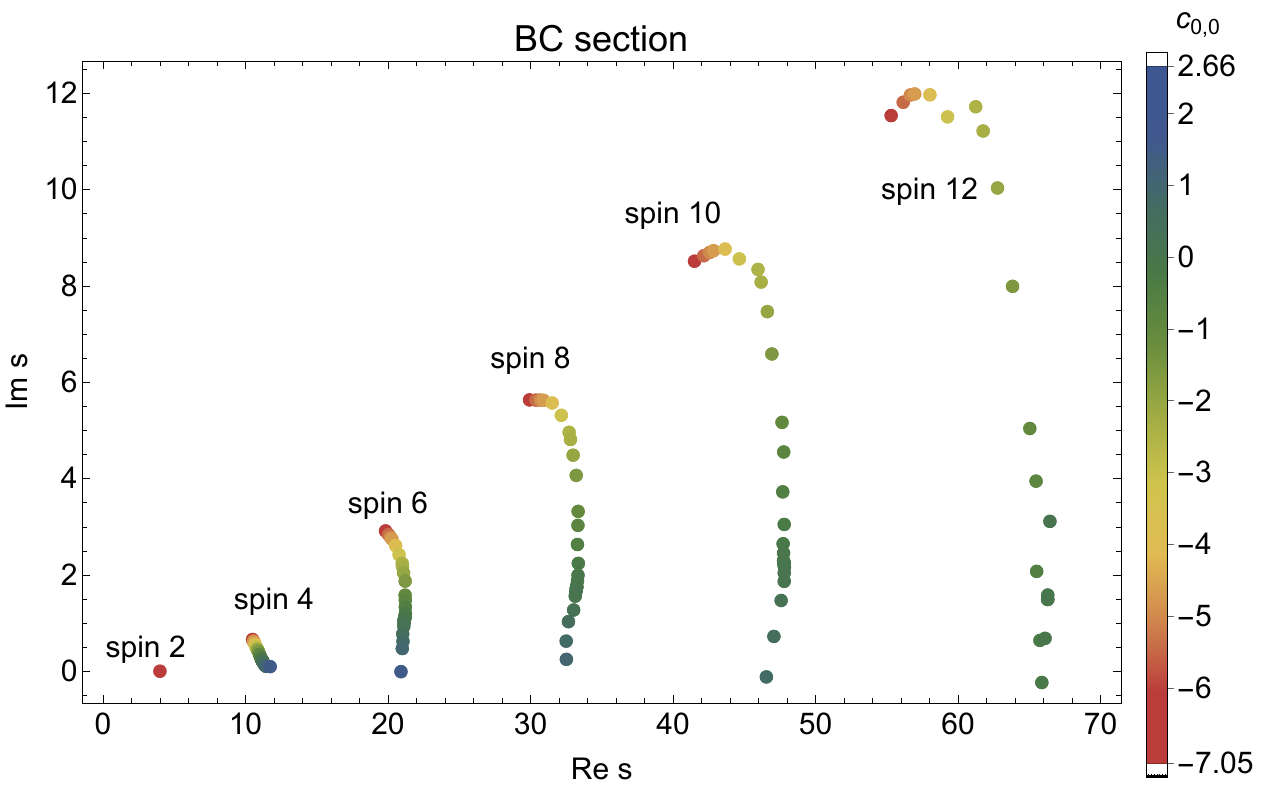}
    \caption{
    Motion of resonance zeros in the complex plane along the lower and upper arcs. The comparison illustrates the difference between energy decoupling along $AC$ and fixed-mass large-$N_c$-like decoupling along $BC$.
    }
    \label{fig:resonances_lower_upper}
\end{figure}

Along the $BC$ arc, the relevant masses remain approximately fixed, while residues and effective couplings decrease. This is the large-$N_c$-like decoupling mechanism discussed in the main text. In the right panel of Fig.~\ref{fig:resonances_lower_upper}, we show the motion of the resonances on the leading trajectory in the complex $s$ plane, with the color coding their dependence on $c_{0,0}$. Their real parts remain approximately constant, while their imaginary parts rapidly approach zero. In this regime numerical convergence becomes significantly slower, and we are therefore unable to extract the precise rate of decoupling. Nevertheless, the behavior is consistent with a suppression proportional to $c_{1,0}$, as observed for the spin-two threshold bound state. For comparison, the left panel of Fig.~\ref{fig:resonances_lower_upper} shows the corresponding motion along the $AC$ arc, where the zeros move deeper into the complex plane as $c_{0,0}$ is varied.

\begin{figure}[h!]
    \centering
    \includegraphics[width=0.4\linewidth]{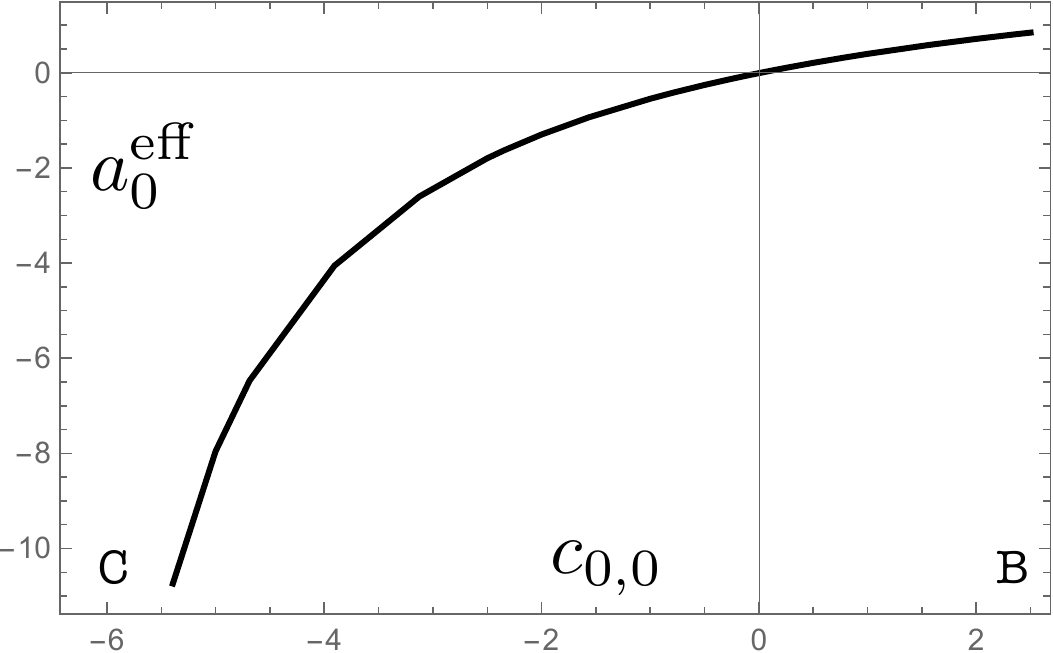}
    \caption{
    Effective scalar threshold parameter along the $BC$ arc. Its behavior tracks the approach of the scalar virtual state to the threshold pole before the zero-pole cancellation near the $C$ cusp.
    }
    \label{fig:effective_scatt_length_BC}
\end{figure}

The scalar channel also exhibits the zero-pole collision described in Appendix~\ref{app:cdd}. It is useful to track the effective spin-zero scattering length $a_0^{\rm eff}$ defined in Appendix~\ref{sec:K-matrix}; its behavior is shown in Fig.~\ref{fig:effective_scatt_length_BC}. It mirrors the scattering-length dynamics on the lower branch, but in reverse. It attains its maximum at the $B$ cusp, vanishes at $c_{0,0}=0$ where the spin-zero virtual state emerges, and then diverges to negative values as the virtual state approaches threshold and collides with the pole.

\begin{figure}[h!]
    \centering
    \includegraphics[width=0.45\linewidth]{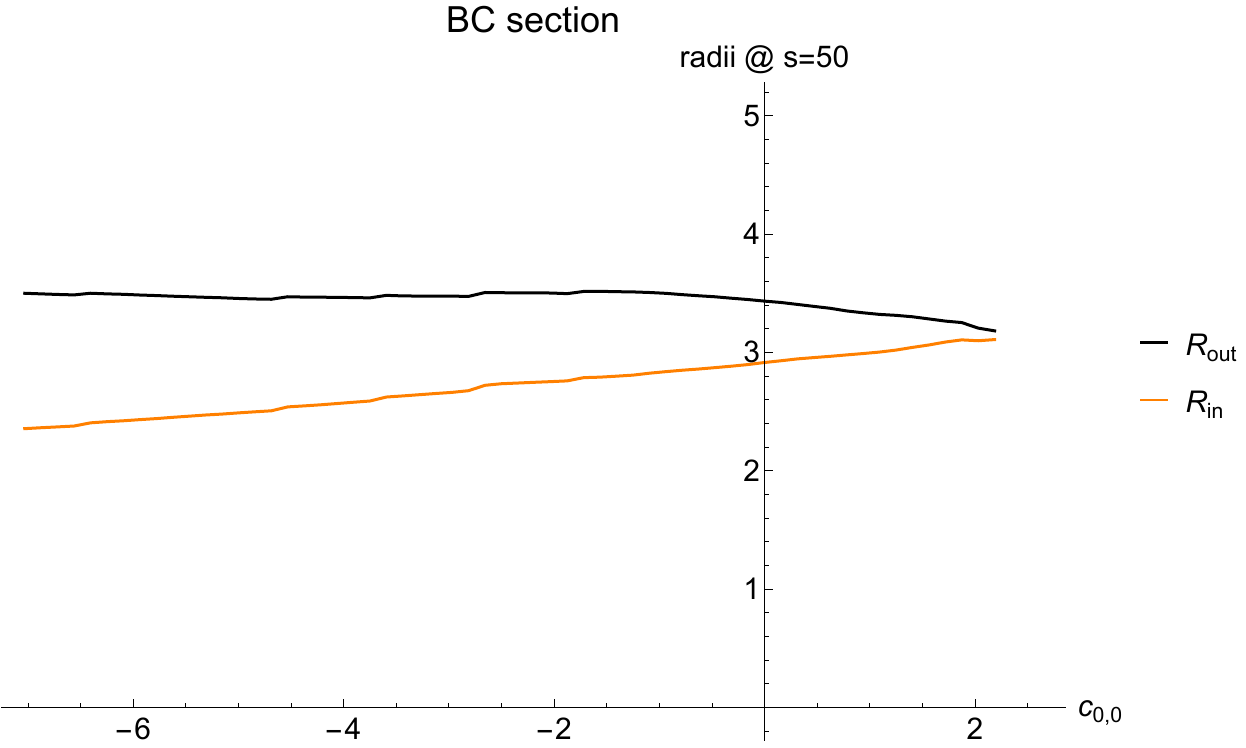}
    \caption{
    Diffractive structure of the amplitude along the $BC$ arc. The observed sharpening is consistent with the narrowing of resonances at approximately fixed mass.
    }
    \label{fig:diff_cone_BC}
\end{figure}

Finally, along the $BC$ arc we observe the same impact-parameter profile found in Ref.~\cite{Correia:2025uvc}, characterized by a ring structure. In Fig.~\ref{fig:diff_cone_BC}, we plot the inner and outer radii of the ring, extracted by fitting the diffractive-cone profile at fixed $s=50$. Along this branch it is meaningful to compare these radii as functions of $c_{0,0}$, since the spectrum remains approximately stable. We observe an interesting shrinkage of the ring as we move toward the large-$N_c$-like decoupling regime near the $B$ cusp. It would be interesting to understand whether the rate of shrinkage can be related quantitatively to the decoupling of the higher-spin sector.

\section{$\lambda\phi^4$ at one loop}
\label{app:one-loop}

Here we review the formulas used for the perturbative $\lambda\phi^4$ analysis. At tree level, $T(s,t)=-\lambda+O(\lambda^2)$. At one loop there is no particle production, and elastic unitarity in the $s$-channel fixes the leading imaginary part,
\begin{equation}
\Im T(s,t)
=
\begin{cases}
\displaystyle
\frac{1}{2}\frac{\lambda^2}{16\pi}
\frac{\sqrt{s-4}}{\sqrt{s}},
& s\geq4, \\[0.8em]
0,
& -t<s<4 .
\end{cases}
\label{eq:im}
\end{equation}
Inserting this discontinuity into the fixed-$t$ dispersion relation gives
\begin{equation}
T(s,t)
=
-\lambda
+
\lambda^2 \mathcal I^1(s,t)
+
O(\lambda^3),
\qquad
\mathcal I^1(s,t)
=
\frac{1}{16\pi^2}
\left[
F(s)+F(t)+F(4-s-t)
\right],
\label{eq:re}
\end{equation}
where
$F(s)=\sqrt{2}\arctan(\frac{1}{\sqrt{2}})
-\frac{\sqrt{4-s}}{\sqrt{s}}\arctan\frac{\sqrt{s}}{\sqrt{4-s}}$.
Note that, With the standard analytic continuation above the cut, $\Im F(s)$ for $s>4$ reproduces~\eqref{eq:im}.

For $0<s<4$, the partial-wave projections are real. Using the definitions in App.~\ref{sec:numerics}
\begin{equation}
f_\ell(s)
=
-\frac{\lambda}{16\pi}\delta_{\ell0}
+
\lambda^2\mathcal I_\ell^1(s)
+
O(\lambda^3),
\qquad
\mathcal I_\ell^1(s)
=
\frac{1}{32\pi}
\int_{-1}^{1} dz\,
P_\ell(z)\,
\mathcal I^1(s,t(z)).
\label{eq:Iell_direct}
\end{equation}
The tree-level term contributes only to spin zero. For even $\ell\geq2$, the leading contribution is therefore one loop and $O(\lambda^2)$.

It is often useful to rewrite Eq.~\eqref{eq:Iell_direct} directly in dispersive form. Setting $a=(4-s)/2$, one finds, for even $\ell\geq2$,
\begin{equation}
\mathcal I_\ell^1(s)
=
\frac{1}{256\pi^3 a}
\int_4^\infty dv\,
\frac{\sqrt{v-4}}{\sqrt v}\,
Q_\ell\left(\frac{v-a}{a}\right).
\label{eq:Iell_phi4_disp}
\end{equation}
Odd spins vanish by crossing symmetry. This form makes the higher-spin virtual-state condition transparent. Below threshold, $\rho(s)=i\sqrt{4-s}/\sqrt{s}$, and $S_\ell(s)=1+i\rho(s)f_\ell(s)=0$ becomes
\begin{equation}
1
-
\frac{\sqrt{4-s}}{\sqrt{s}}\,
\lambda^2\mathcal I_\ell^1(s)
=
0,
\qquad
\ell\geq2 .
\label{eq:higher_spin_virtual_phi4}
\end{equation}
The higher-spin virtual-state condition therefore depends only on $\lambda^2$, and is insensitive to the sign of the quartic coupling.

We now match $\lambda$ to the low-energy coordinate $c_{0,0}$. By definition, $T/(32\pi)=c_{0,0}+\cdots$ at the crossing-symmetric point $s=t=u=4/3$. Let $F_*=F(4/3)$. Evaluating Eq.~\eqref{eq:re} at this point gives
\begin{equation}
32\pi c_{0,0}
=
-\lambda
+
\frac{3\lambda^2}{16\pi^2}F_*
+
O(\lambda^3),
\qquad
\lambda
=
-32\pi c_{0,0}
+
192F_*\,c_{0,0}^2
+
O(c_{0,0}^3).
\label{eq:lambda_c00}
\end{equation}
Substituting back, the one-loop amplitude matched at the crossing-symmetric point is
\begin{equation}
\frac{T(s,t)}{32\pi}
=
c_{0,0}
+
\frac{2}{\pi}c_{0,0}^2
\left[
F(s)+F(t)+F(u)-3F_*
\right]
+
O(c_{0,0}^3).
\label{eq:matched_phi4}
\end{equation}

Finally, we recall the running-coupling estimate used in the main text. The scale dependence follows from the subtracted dispersive representation of the one-loop bubble. For a Euclidean scale $\mu$,
\begin{equation}
F(-\mu^2)-F(-\mu_0^2)
=
\frac{1}{\pi}
\int_4^\infty dv\,
\Im F(v)
\left(
\frac{1}{v+\mu^2}
-
\frac{1}{v+\mu_0^2}
\right)
=
-\log\frac{\mu}{\mu_0}
+\cdots .
\end{equation}
Including the three channels gives the usual leading-log running
\begin{equation}
\lambda(\mu)
=
\frac{\lambda_*}
{1-\frac{3\lambda_*}{16\pi^2}
\log\frac{\mu}{\mu_*}},
\qquad
\lambda_*=-32\pi c_{0,0},
\label{eq:running_lambda_phi4}
\end{equation}
where $\mu_*$ denotes the subtraction scale associated with the crossing-symmetric matching point. For the negative quartic coupling relevant on the $AB$ arc, $\lambda_*<0$, and $\lambda(\mu)$ approaches zero logarithmically in the ultraviolet.

Using elastic unitarity with $\mu=\sqrt{s}$ gives
\begin{equation}
\Im T(s,0)
=
\frac{1}{2}
\frac{\lambda^2(\sqrt{s})}{16\pi}
\frac{\sqrt{s-4}}{\sqrt{s}}
+
O(\lambda^3),
\qquad
\sigma_{\rm tot}(s)
=
\frac{\Im T(s,0)}{\sqrt{s(s-4)}} .
\label{eq:running_im_phi4}
\end{equation}
Therefore
\begin{equation}
\sigma_{\rm tot}(s)
=
\frac{1}{32\pi s}
\left[
\frac{\lambda_*}
{1-\frac{3\lambda_*}{16\pi^2}
\log\frac{\sqrt{s}}{\mu_*}}
\right]^2
+
O(\lambda_*^3).
\label{eq:running_sigma_phi4}
\end{equation}
For $\lambda_*<0$, this gives the asymptotic estimate
\begin{equation}
\sigma_{\rm tot}(s)
\underset{s\to\infty}{\sim}
\frac{8\pi^3}{9}\,
\frac{1}{s\log^2(\sqrt{s}/\mu_*)},
\end{equation}
up to subleading logarithmic corrections. This is the running-coupling estimate used for comparison with the numerical cross-sections near the free theory.

\section{A simple CDD model for virtual states and zero-pole cancellation}
\label{app:cdd}

A simple way to model the spin-zero partial wave in the presence of a virtual state is through a (Castillejo-Dalitz-Dyson) CDD factor.
A convenient crossing-symmetric representation is
\begin{equation}
S^{\rm CDD}(s,m_V^2)
=
\frac{\sqrt{m_V^2(4-m_V^2)}-\sqrt{s(4-s)}}
{\sqrt{m_V^2(4-m_V^2)}+\sqrt{s(4-s)}} .
\label{eq:cdd_factor_s}
\end{equation}
This factor is elastic, symmetric under $s\to 4-s$, and has zeros at $s=m_V^2$ and $s=4-m_V^2$. After analytic continuation through the elastic cut, these zeros correspond to second-sheet poles, i.e. virtual states.

With our convention $S_0(s)=1+i\rho(s)f_0(s)$, where $\rho(s)=\sqrt{s-4}/\sqrt{s}$, the corresponding partial wave is
\begin{equation}
f_0^{\rm CDD}(s)
=
\frac{S^{\rm CDD}(s)-1}{i\rho(s)}.
\label{eq:f0_cdd}
\end{equation}
Let the virtual state approach the two-particle threshold as $m_V^2=4-\epsilon^2$, with $\epsilon\ll1$. Expanding near threshold gives
\begin{equation}
\Re f_0^{\rm CDD}(s)
=
-\frac{4}{\epsilon}
+
\frac{4}{\epsilon^3}(s-4)
+\cdots .
\end{equation}
Thus the scattering length, $a=-f_0(4)/2$, diverges as $a\simeq2/\epsilon$, or equivalently
\begin{equation}
4-m_V^2=\epsilon^2\simeq \frac{4}{a^2}.
\label{eq:virtual_scattering_length}
\end{equation}
This is the standard universal relation between the distance from threshold and the scattering length. The CDD model therefore provides an explicit realization of the threshold behavior described by the $K$-matrix analysis of Appendix~\ref{sec:K-matrix}.

We now use the same language to model the disappearance of the scalar threshold state along the $BC$ arc. Near threshold, write $p(s)=\sqrt{s-4}/2$. A unitary CDD factor containing a bound-state pole at $p=i\kappa_B$ and a zero at $p=i\kappa_Z$ can be written as
\begin{equation}
S_0(p)
=
\frac{p+i\kappa_B}{p-i\kappa_B}
\frac{p-i\kappa_Z}{p+i\kappa_Z}.
\label{eq:cdd_zero_pole}
\end{equation}
The first factor contains the pole, while the second contains the zero. When the zero collides with the pole, $\kappa_Z\to\kappa_B$, the two factors cancel and $S_0(p)\to1$. Since $f_0=(S_0-1)/(i\rho)$, this cancellation removes the threshold singularity associated with the scalar bound state.

This gives a local model for the mechanism observed on the $BC$ arc. The scalar threshold pole present near the $B$ cusp is accompanied by a virtual-state zero. As one moves toward the $C$ cusp, the zero approaches the threshold pole and cancels it. After the collision the scalar partial wave is regular at threshold, providing a natural continuation to the $AC$ arc, where the repulsive scalar interaction no longer supports a spin-zero virtual state.
\section{Threshold behavior from the $K$-matrix}
\label{sec:K-matrix}

In this appendix we derive the threshold behavior of a partial wave when a bound state of spin $\ell$ sits at the two-particle threshold. We use the convention
\begin{equation}
    S_\ell(s)=1+i\rho(s)f_\ell(s),
    \qquad
    \rho(s)=\frac{\sqrt{s-4}}{\sqrt{s}},
\end{equation}
so elastic unitarity implies $\Im f_\ell^{-1}(s)=-\rho(s)/2$. 
Writing $p(s)=\sqrt{s-4}/2$, we use the standard $K$-matrix parametrization, see e.g.~\cite{Correia:2025ozf},
\begin{equation}
    f_\ell(s)=
    \frac{p^{2\ell}(s)}
    {K_\ell^{-1}(s)-\frac{i}{2}\rho(s)p^{2\ell}(s)} .
\end{equation}
A bound state below threshold, $s_B<4$, appears as a pole of the partial wave. It therefore satisfies
\begin{equation}
    K_\ell^{-1}(s_B)
    =
    \frac{i}{2}\rho(s_B)p^{2\ell}(s_B),
\end{equation}
where below threshold $\rho(s_B)=i\sqrt{4-s_B}/\sqrt{s_B}$. As $s_B\to4$, the right-hand side vanishes, and hence a threshold bound state implies $K_\ell^{-1}(4)=0$.

Expanding $K_\ell^{-1}(s)=\alpha_\ell p^2(s)+\cdots$, we find, for even $\ell\geq2$,
\begin{equation}
    f_\ell(s)
    \simeq
    \frac{p^{2\ell-2}(s)}{\alpha_\ell}
    =
    \frac{(s-4)^{\ell-1}}{4^{\ell-1}\alpha_\ell}.
\end{equation}
This is the threshold behavior used in the main text. We define the effective threshold coupling by
\begin{equation}
    f_\ell(s)\sim -g_{\rm eff}^2(s-4)^{\ell-1},
    \qquad \ell\geq2,
\end{equation}
or equivalently $g_{\rm eff}^2=-1/(4^{\ell-1}\alpha_\ell)$. With this convention the threshold bound states observed on the boundary have positive $g_{\rm eff}^2$. By contrast, an ordinary finite threshold parameter gives $f_\ell(s)\sim a_\ell(s-4)^\ell$, with $a_\ell\geq0$; the threshold bound state is therefore signalled by the reduced power and sign.

It remains useful to match this normalization to a pole model. A spin-$\ell$ pole contribution with the angular-momentum barrier included,
$T_\ell(s,t)\simeq g^2(s-4)^\ell P_\ell(\cos\theta)/(s_B-s)$, projects onto
\begin{equation}
    f_\ell(s)
    \simeq
    \frac{g^2(s-4)^\ell}
    {16\pi(2\ell+1)(s_B-s)}
    \xrightarrow{s_B\to4}
    -\frac{g^2}{16\pi(2\ell+1)}(s-4)^{\ell-1}.
\end{equation}
Thus, for the spin-two threshold state,
$f_2(s)\simeq -g^2(s-4)/(80\pi)$, so that
$g_{\rm eff}^2=g^2/(80\pi)$ in the normalization used in Fig.~\ref{fig:bcfigs}.

The scalar channel is special. For $\ell=0$, the $K$-matrix form reduces to $f_0(s)=1/(K_0^{-1}(s)-i\rho(s)/2)$. A threshold bound state implies $K_0^{-1}(4)=0$, and the leading behavior is the universal unitarity singularity $f_0(s)\sim2i/\rho(s)$. Thus no finite scalar threshold residue can be defined. One may nevertheless characterize the scalar threshold state through the subleading finite part
\begin{equation}
    a_0^{\rm eff}
    =
    \lim_{s\to4^+}
    \left[
    f_0(s)-\frac{2i}{\rho(s)}
    \right],
\end{equation}
which provides a useful diagnostic of virtual states approaching threshold.

\section{Exploring further directions of the scalar island}
\label{sec:3dshape}

In this section we turn on an additional direction in the space of Wilson coefficients and study the primal and dual projections of the scalar island. We present the island in the three-dimensional space $\{c_{0,0},c_{1,0},c_{0,1}\}$.
The last coordinate admits the following dispersive representation:
\bea
c_{0,1} &= \frac{1}{\pi} \int_4^\infty \!\!\!\!\! dv \, \frac{3 \, \Im T
(v,\frac{4}{3})}{(v-\frac{4}{3})^4} - \frac{2 \, \partial_t \Im M(v,t)|_{t=\frac{4}{3}}}{(v-\frac{4}{3})^3}, \label{eq:c3}
\eea
where both terms in the integrand are sign-definite, and the second term receives contributions from all spins except the scalar.

The amplitudes we have characterized in this Letter are special from this three-dimensional point of view, as they correspond to edges of the $(c_{0,0},c_{1,0},c_{0,1})$ space. 

\begin{table}[h!]
\centering
\begin{tabular}{c|ccc}
\hline
Point & $c_{0,0}$ & $c_{1,0}$ & $c_{0,1}$ \\
\hline
Cusp $A$ & $0$ & $0$ & $0$ \\
Cusp $B$ & $2.66$ & $0.0618$ & $0.0587$ \\
Cusp $C$ & $-7.06$ & $0.930$ & $-1.81$ \\
Prevertex $D$ & $-1.41$ & $0.034$ & $-0.009$ \\
\hline
\end{tabular}
\caption{Coordinates of the special points in the $(c_{0,0},c_{1,0},c_{0,1})$ space from the best primal numerics. }
\label{tab:cusp_points}
\end{table}

\begin{figure}[t!]
    \centering
    \includegraphics[scale=0.5]{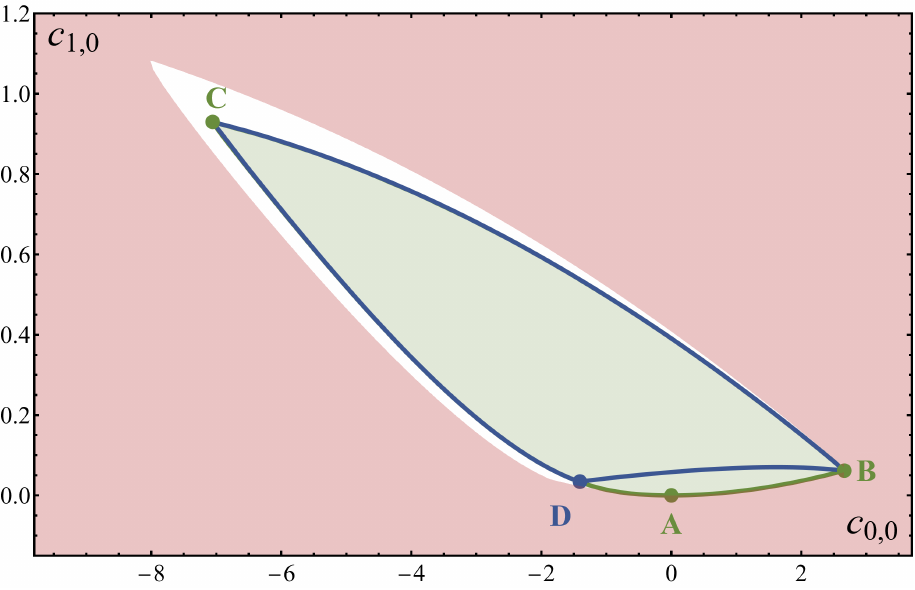}
    \includegraphics[scale=0.5]{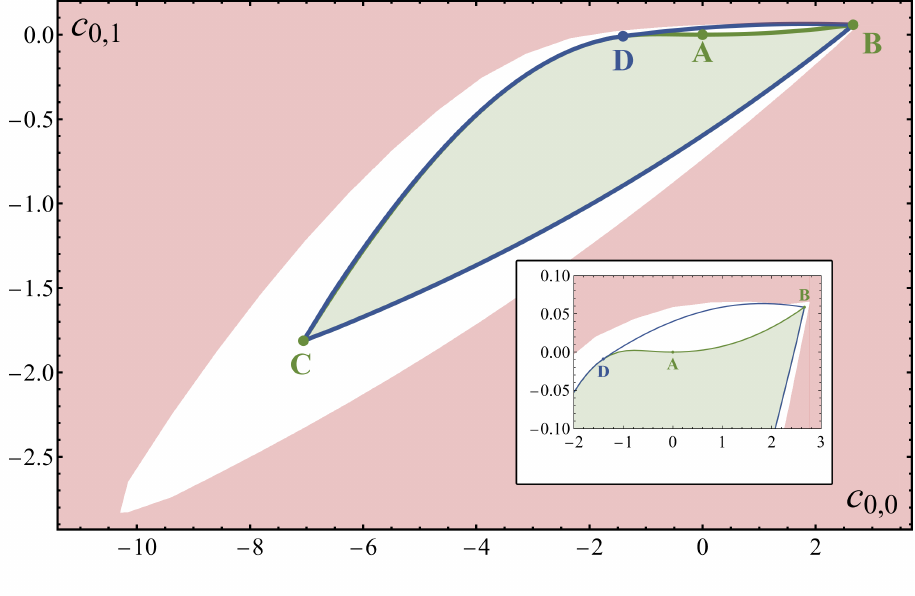} \\
    \includegraphics[scale=0.4]{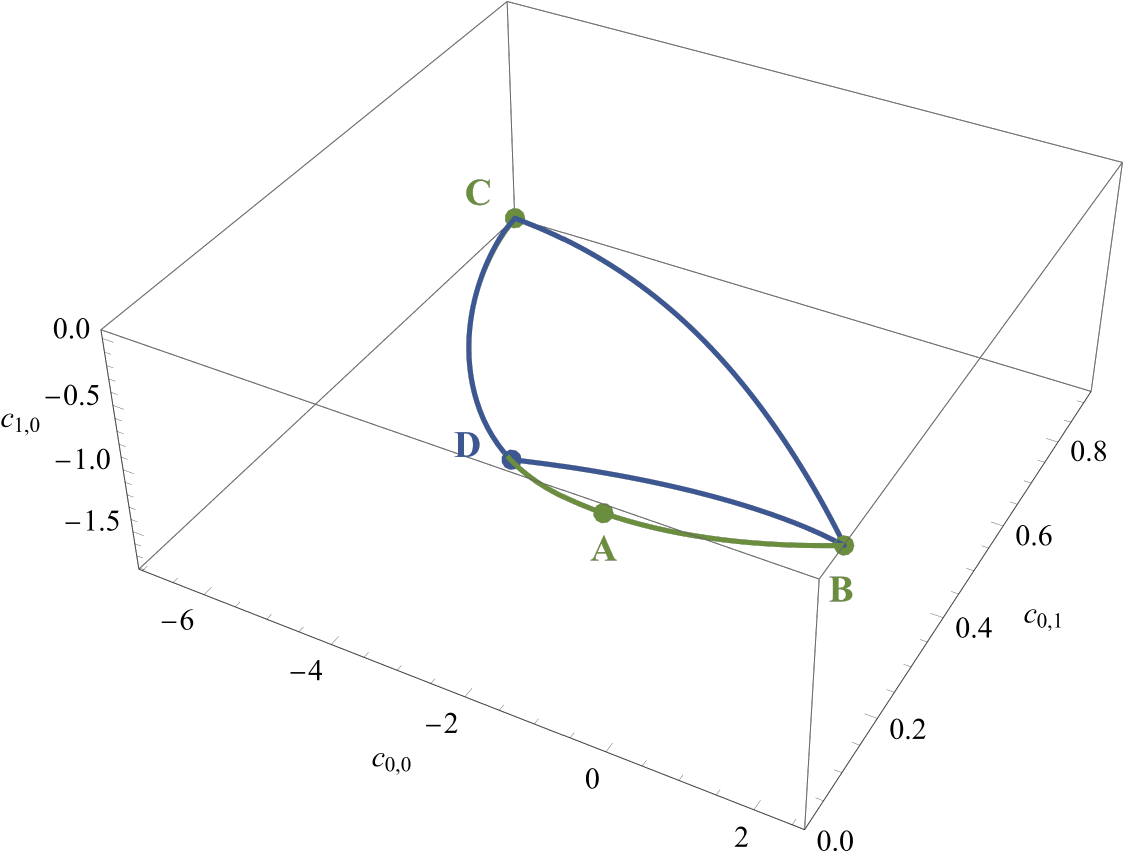}
    \caption{Various projections of the four-dimensional scalar S-matrix space in the $(c_{0,0},c_{1,0},c_{0,1})$ space. Blue solid/dashed curves show the extremal amplitudes in the $(c_{0,0},c_{0,1})$ and $(c_{0,1},c_{1,0})$ planes, respectively.}
    \label{fig:c0c2space}
\end{figure}

It would be interesting to understand which quantity develops a discontinuity when crossing one of these edges. There is an additional edge connecting the $B$ cusp to a point along the $AC$ arc that does not exhibit any obvious discontinuous feature. This is a \emph{prevertex}, a structure previously observed in the monolith of $O(N)$ theories in $1+1$ dimensions \cite{Cordova:2019lot}, and here identified for the first time in higher dimensions. We were not able to correlate the prevertex $D$ with any specific feature of the amplitudes on the $AC$ arc. The only mild correlation we observe is with the minimum of the homogeneous ratios of S-matrix data $g_3$ and $g_4$ in Appendix~\ref{app:low_energy_constants}, although we have not found a quantitative explanation for this.

An interesting property of the $AD$ and $DB$ arcs is that they interchange their roles when projecting onto different faces of the three-dimensional island. In particular, the $BD$ edge is extremal in the $(c_{0,0},c_{0,1})$ plane. Moreover, we observe that amplitudes with $c_{0,1}<0$ are always associated with Regge trajectories satisfying $\alpha_\text{eff}(0) \simeq 1$. This suggests that the $BD$ edge separates two distinct surfaces with different Regge behavior.

Interestingly, the $BD$ edge provides an alternative mechanism for large-$N_c$-like decoupling compared to the one observed along the $BC$ edge. It interpolates between a theory with a strongly coupled higher-spin sector at moderately high energy (recall that the mass of the lightest spin-two resonance at that point is $M_2\sim 3.4$ and the theory at the $B$ cusp, which has no resonances. The amplitudes along this edge may therefore provide a more realistic candidate for large-$N_c$ $\eta^\prime$ scattering, potentially closer to QCD. At the $BC$ cusp, the presence of threshold bound states in both spin zero and spin two leads, in the large-$N_c$ decoupling limit, to an almost meromorphic amplitude with a strongly coupled S-wave and a D-wave with cutoff at 4. Along the $BD$ edge, on the other hand, we expect a nontrivial S-wave dynamics changing from weakly repulsive at $D$ to highly attractive interaction at $B$, while the D-wave with a cutoff stays of order a few times the mass of the $\eta^\prime$.

\twocolumngrid

\end{document}